\documentclass[a4paper,fleqn,usenatbib]{mnras}
\usepackage[T1]{fontenc}
\usepackage{ae,aecompl}
\usepackage{graphicx}
\title[3D structure of the SMC]{The VMC Survey -- XL. Three-dimensional
structure of the Small Magellanic Cloud as derived from red clump stars}
\author[Tatton et al.]{
B.L. Tatton,$^{1}$
J.Th. van Loon,$^{1}$
M.-R.L. Cioni,$^{2}$
K. Bekki,$^{3}$
C.P.M. Bell,$^{2}$
\newauthor
S. Choudhury,$^{4,5}$
R. de Grijs,$^{4,5,6}$
M.A.T. Groenewegen,$^{7}$
V.D. Ivanov,$^{8}$
\newauthor
M. Marconi,$^{9}$
J.M. Oliveira,$^{1}$
V. Ripepi,$^{9}$
S. Rubele,$^{10,11}$
S. Subramanian$^{12}$
\newauthor
and N.-C. Sun$^{13}$\\
$^{1}$  Lennard-Jones Laboratories, Keele University, ST5 5BG, UK\\
$^{2}$  Leibniz-Institut f\"ur Astrophysik Potsdam (AIP), An der Sternwarte
       16, D-14482 Potsdam, Germany\\
$^{3}$  ICRAR, M468, The University of Western Australia, 35 Stirling Hwy,
       Crawley WA 6009, Australia\\
$^{4}$  Department of Physics and Astronomy, Macquarie University, Balaclava
       Road, NSW 2109, Australia\\
$^{5}$  Research Centre for Astronomy, Astrophysics and Astrophotonics,
       Macquarie University, Balaclava Road, Sydney, NSW 2109, Australia\\
$^{6}$  International Space Science Institute--Beijing, 1 Nanertiao,
       Zhongguancun, Hai Dian District, Beijing 100190, China\\
$^{7}$  Koninklijke Sterrenwacht van Belgi\"e, Ringlaan 3, B-1180 Brussels,
       Belgium\\
$^{8}$  European Southern Observatory, Karl-Schwarzschild-Stra{\ss}e 2,
       D-85748 Garching bei M\"unchen, Germany\\
$^{9}$  INAF, Osservatorio Astronomico di Capodimonte, Via Moiarello 16,
       I-80131 Napoli, Italy\\
$^{10}$ Dipartimento di Fisica e Astronomia, Universit\`a di Padova, Vicolo
       dell’Osservatorio 2, I-35122 Padova, Italy\\
$^{11}$ INAF, Osservatorio Astronomico di Padova, Vicolo dell’Osservatorio 5,
       I-35122 Padova, Italy\\
$^{12}$ Indian Institute of Astrophysics, II Block, Koramangala, Bengaluru 560
       034, India\\
$^{13}$ Department of Physics and Astronomy, The University of Sheffield, Hicks
       Building, Hounsfield Road, Sheffield S3 7RH, UK}
\date{Submitted: 13 May 2020; Resubmitted: 30 September 2020}
\pubyear{2020}
\begin{document}
\label{firstpage}
\pagerange{\pageref{firstpage}--\pageref{lastpage}}
\maketitle
\begin{abstract}
Galaxy interactions distort the distribution of baryonic matter and can affect
star formation. The nearby Magellanic Clouds are a prime example of an ongoing
galaxy interaction process. Here we use the intermediate-age ($\sim1$--$10$
Gyr) red clump stars to map the three-dimensional structure of the Small
Magellanic Cloud (SMC) and interpret it within the context of its history of
interaction with the Large Magellanic Cloud (LMC) and the Milky Way. Red clump
stars are selected from near-infrared colour--magnitude diagrams based on data
from the VISTA survey of the Magellanic Clouds. Interstellar reddening is
measured and removed, and the corrected brightness is converted to a distance,
on a star-by-star basis. A flat plane fitted to the spatial distribution of
red clump stars has an inclination $i=35\degr$--$48\degr$ and position angle
PA$=170\degr$--$186\degr$. However, significant deviations from this plane are
seen, especially in the periphery and on the eastern side of the SMC. In the
latter part, two distinct populations are present, separated in distance by as
much as 10 kpc. Distant red clump stars are seen in the North of the SMC, and
possibly also in the far West; these might be associated with the predicted
`Counter-Bridge'. We also present a dust reddening map, which shows that dust
generally traces stellar mass. The structure of the intermediate-age stellar
component of the SMC bears the imprints of strong interaction with the LMC a
few Gyr ago, which cannot be purely tidal but must have involved ram pressure
stripping.
\end{abstract}
\section{Introduction}

The details of how galaxies have evolved over cosmological times are imprinted
in their star formation history (SFH), chemical enrichment, and morphological
and kinematic structure. Many galaxies have experienced interactions with
other galaxies, and this will have affected their evolution. If we are to
understand galaxy evolution, we need to understand what happens during galaxy
interactions. A prime example of a galaxy interaction is found in our backyard
-- the Magellanic System, comprising the Large Magellanic Cloud (LMC) and the
Small Magellanic Cloud (SMC) at distances of just $\approx 50$ kpc
\citep[LMC;][]{degrijs14} and $\approx 61$ kpc \citep[SMC;][]{degrijs15}, the
Magellanic Bridge between them, the purely gaseous Magellanic Stream trailing
them, and the Leading Arm ahead of them in their flight through the Milky Way
halo. In this paper we investigate the structure of the SMC as seen in its
intermediate-age ($\sim1$--$10$ Gyr) red-clump stars and link it to the recent
LMC--SMC--Milky Way interaction history.

While their three-dimensional (3-D) space motions have been used to argue that
the Magellanic Clouds are on their first close passage of the Milky Way, this
is not certain and depends on the Milky Way's mass
\citep{kallivayalil13,patel17}. In any case, mutual interactions between the
LMC and SMC may have occurred more often and over a longer period of time
\citep{besla16}. Gradients in gravity and momentum have led to tidal
stretching and bending, most notably in the form of an eastern extension to
the SMC known as the `Wing' \citep{shapley40,gonidakis09,belokurov19}; this is
now also seen in the internal kinematics \citep[e.g.,][]{niederhofer18}. While
the Bridge is for certain a result of the galaxy interactions, too, it has
proven to be a more complex structure than initially thought, with different
distributions of gas \citep{hindman63b}, young stars \citep{irwin90} and old
stars \citep{belokurov17}. The existence of the Magellanic Stream as well as
recent proper motion measurements of stars within the SMC both indicate that a
physical collision has happened between the LMC and SMC $\sim200$ Myr ago
\citep{hammer15,zivick18}.

\begin{table*}
\caption{Line-of-sight depths for the SMC from the literature.}
\begin{tabular}{lcr}
\hline\hline
\noalign{\smallskip}
Publication & Depth (kpc) & Notes \\
\noalign{\smallskip}
\hline
\noalign{\smallskip}
\citet{hatzidimitriou89} & 17, 7         & $2\sigma$ North--East and South--West \\
\citet{gardiner91}       & 4--16         & $2\sigma$ depth dependent on portion observed (North and North--West) \\
\citet{groenewegen00}    & 14            & (OGLE-II, DENIS and 2MASS) Cepheids, front--back \\
\citet{crowl01}          & $(4.1\pm0.8)$--$(6.3\pm1.3)$ & Cepheids \\
\citet{glatt08}          & $\sim10$      & deep Hubble Space Telescope imaging \\
\citet{subramanian09}    & $3.4\pm1.2$   & OGLE-II RC: average depth of the disc \\
\citet{subramanian09}    & $4.2\pm1.0$   & OGLE-II RC: average depth in the North \\
\citet{subramanian09}    & $2.6\pm0.8$   & OGLE-II RC: average depth in the South \\
\citet{subramanian09}    & $3.1\pm1.0$   & OGLE-II RC: average depth in the East \\
\citet{subramanian09}    & $2.8\pm0.9$   & OGLE-II RC: average depth in the West \\
\citet{kapakos11}        & $4.13\pm0.27$ & OGLE-II/III RR\,Lyr{\ae} {\it V}-band: Central area $1\sigma$ \\
\citet{haschke12c}       & $4.2\pm0.4$   & OGLE-III RR\,Lyr{\ae} \\
\citet{haschke12c}       & 5.14--6.02    & OGLE-III Cepheids; dependent on field selection \\
\citet{kapakos12}        & $5.3\pm0.4$   & OGLE-II/III RR\,Lyr{\ae} {\it V}-band: Extended region $1\sigma$ \\
\citet{subramanian12}    & $4.57\pm1.03$ & $1\sigma$ OGLE-III RC stars \\
\citet{subramanian12}    & 6--8          & $1\sigma$ OGLE-III RC stars (North--Eastern regions) \\
\citet{subramanian12}    & $\sim14$      & $4\sigma$ OGLE-III RR\,Lyr{\ae} \& RC stars front to back \\
\citet{subramanian15}    & $8.1\pm1.4$   & OGLE-III SMC disc $z$-distribution \\
\citet{subramanian15}    & $1.8\pm0.6$   & OGLE-III SMC disc orientation corrected \\
\citet{deb15}            & $4.9\pm0.7$   & OGLE-III RR\,Lyr{\ae} (RRab); Fourier relation \\
\citet{deb15}            & $4.1\pm0.7$   & OGLE-III RR\,Lyr{\ae} (RRab); metallicity relation \\
\noalign{\smallskip}
\hline
\end{tabular}
\label{tab:smclos}
\end{table*}

\begin{table*}
\caption{Position angle (PA) and inclination ($i$) for the SMC from the literature.}
\begin{tabular}{lccr}
\hline\hline
\noalign{\smallskip}
Publication            & $i$ ($\degr$)   & PA ($\degr$)   & Notes \\
\noalign{\smallskip}
\hline
\noalign{\smallskip}
\citet{caldwell86}     & $70\pm3$        & $58\pm10$      & Cepheids \\
\citet{laney86}        & $45\pm7$        & $55\pm17$      & Cepheids \\
\citet{groenewegen00}  & $68\pm2$        & $238\pm7$*     & (OGLE-II, 2MASS and DENIS) Cepheids \\
\citet{kunkel00}       & $73\pm4$        &                & Radial velocities Carbon stars \\
\citet{stanimirovic04} & $40\pm20$       & $40\pm10$      & H\,{\sc i} \\
\citet{haschke12c}     & $74\pm9$        & $66\pm15$      & OGLE-III Cepheids \\
\citet{haschke12c}     & $7\pm15$        & $83\pm21$      & OGLE-III RR\,Lyr{\ae} \\
\citet{subramanian12}  & 0.58            & 55.5           & OGLE-III RC axes ratio \\
\citet{subramanian12}  & 0.50            & 58.3           & OGLE-III RR\,Lyr{\ae} axis \\
\citet{dobbie14a}      & 25--70          & 120--130       & Radial velocities RGB stars \\
\citet{deb15}          & $2.3\pm0.8$     & $74.3\pm0.5$   & OGLE-III RRab stars \\
\citet{deb15}          & $0.51\pm0.29$   & $56.0\pm0.8$   & OGLE-III Main body \\
\citet{deb15}          & $2.244\pm0.024$ & $85.54\pm0.33$ & OGLE-III North--Eastern arm \\
\citet{subramanian15}  & $63.1\pm1.0$    & $152\pm8$      & OGLE-III Cepheids fundamental-mode \\
\citet{subramanian15}  & $66.3\pm0.9$    & $159\pm10$     & OGLE-III Cepheids first-overtone \\
\citet{subramanian15}  & $64.4\pm0.7$    & $155\pm6$      & OGLE-III Cepheids combined \\
\citet{rubele15}       & $39.3\pm5.5$    & $179.3\pm2.1$  & CMDs \\
\citet{deb17}          & $2.30\pm0.14$   & $38.5\pm0.3$   & OGLE-IV RRab stars \\
\citet{deb17}          & $3.16\pm0.19$   & $38.0\pm0.6$   & OGLE-IV RRab stars \\
\noalign{\smallskip}
\hline
\noalign{\smallskip}
\multicolumn{4}{l}{*Subtracting $180\degr$ from this PA brings it in line with
the other results.}\\
\end{tabular}
\label{tab:smcpai}
\end{table*}

There exists a rich history of line-of-sight depth measurements in the SMC,
motivated by large values of 20--30 kpc in the eastern part inferred from
Cepheid variables distance measurements by \citet{mathewson86,mathewson88}
\citep[cf.][]{hatzidimitriou93}. \citet{stanimirovic04} reviewed these
findings \citep[see also the more recent review by][]{degrijs15}, noting that
\citet{zaritsky02} had mentioned that differential interstellar extinction can
lead to Cepheid distances being overestimated. Correction for this would bring
most values down to within the tidal radius (4--9 kpc). Even at the time
\citet{welch87} pointed out flaws with these measurements. Further
line-of-sight depth determinations are listed in Table~\ref{tab:smclos}, where
some \citep[e.g.,][]{haschke12c} have interpreted the \citet{crowl01} data as
a 6--12 kpc depth. \citet{kapakos11} argued that the metal-rich stars form a
thin disc-like structure whilst the metal-poor stars form a halo- or
bulge-like structure, appearing thicker in the North--East \citep[see
also][]{kapakos12}.

Measurements of the orientation of the SMC are challenging due to the
stretched nature of the Wing, an elongated young main body and the ellipsoidal
shape of the old population. This is reflected in a large variance in position
angle (PA) and inclination ($i$) values (see Table~\ref{tab:smcpai}). A clear
age dichotomy is seen, with young stellar populations ($<1$ Gyr; Cepheids)
steeply inclined with respect to the plane of the sky while old populations
($>10$ Gyr; RR\,Lyr{\ae}) are shallower. \citet{haschke12c} suggested that
this difference is due to interactions with the LMC and Milky Way. This must
mean that ram pressure on the gas played a role in the interactions, as tidal
forces alone would not differentiate between stars of different ages (nor
between gas and stars). Alternatively, a dwarf--dwarf galaxy merger early on
in the evolution of the SMC could have resulted in a spheroidal old component
and a rotating gaseous disc that has since been truncated by interactions with
the LMC \citep{bekki08}. The merger scenario would predict an intermediate-age
component to be disc like. Furthermore, the main body of the SMC is less
inclined than its periphery; this has been explained as due to tidal forces
between the LMC and SMC \citep{deb15}. The range of position angles is much
smaller in the main body of the SMC than in other regions and this variance
suggests the SMC is not a flat disc.

In this work, we use red clump (RC) stars as tracers of the 3-D structure of
the SMC. RC stars form a metal-rich, younger counterpart to the horizontal
branch. Stars in this phase of stellar evolution are of low mass ($<2.2$
M$_\odot$) and undergo core helium and shell hydrogen burning, starting at an
almost-fixed core mass and hence luminosity \citep{girardi16}. Following the
first ascent of the red giant branch (RGB) the RC phase typically lasts
$\sim0.1$ Gyr (up to 0.2 Gyr). As a result RC stars are easily noticeable in
colour--magnitude diagrams (CMDs) as a clump of stars close to the RGB
\citep[their Figure 4]{salaris12}. As standard candles and more numerous than
variable stars such as RR\,Lyr{\ae} and Cepheids, RC stars are choice tracers
of intermediate-age (a few Gyr) stellar populations in galaxies with
recent/ongoing star formation \citep{girardi01}. Their ages imply they have
had sufficient time to mix well dynamically as, in the absence of strong
rotational support or external influences, an SMC-like dwarf galaxy would
virialise within a few crossing times (well within a Gyr) into a spheroidal
configuration; morphological and kinematical distortions must in general be
due to tidal forces resulting from galaxy interactions. Comparisons with maps
obtained using younger Cepheids \citep{dobrzeniecka16,scowcroft16,ripepi17}
and older RR\,Lyr{\ae} \citep{dobrzeniecka17,muraveva18} can then place these
results in a galactic evolutionary context, as was done for the 2-D case by
\citet{cioni00}, \citet{zaritsky00} and \citet{sun18}.

The studies of SMC depth and structure in general, are hampered by attenuation
of starlight by dust in the interstellar medium (ISM) and its wavelength
dependence \citep{gieren13}. The effect is smaller in the infrared (IR) than
in the optical and we therefore chose to make use of near-IR data. Still, it
can be significant relative to the accuracy we aim to achieve ($\Delta
m\sim0.03$ mag) to trace small differences in distance ($\Delta d<1$ kpc). We
therefore determine the amount of reddening and attenuation and correct for
it. As a by-product, we obtain a map of the interstellar dust distribution,
which traces the ISM and can be compared with the stellar distribution -- see
\citet{tatton13} for an application to the area around 30\,Doradus in the LMC.

The structure of this paper is as follows. Section~\ref{smc:data} describes
the data used; Section~\ref{smc:meth} describes the RC star selection and
de-reddening procedures; Section~\ref{smc:red} presents the reddening map;
Section~\ref{smc:3d} presents the derived 3-D structure; we discuss the
results in Section~\ref{smc:dis} and conclude the paper in
Section~\ref{smc:conc}.

\section{Data}\label{smc:data}

We used near-IR $Y$ and $K_{\rm s}$ photometry from images taken with the
Visible and Infrared Survey Telescope for Astronomy
\citep[VISTA;][]{sutherland15} as part of the VISTA Survey of the Magellanic
Clouds \citep[VMC;][]{cioni11}. These data were obtained with the VISTA
Infrared Camera \citep[VIRCAM;][]{sutherland15} and processed with the VISTA
Data Flow System \citep[VDFS;][]{irwin04,gonzalezfernandez18}. The exposure
times were 40 min.\ for the $Y$ band and 150 min.\ for the $K_{\rm s}$ band.
The 50\% completeness level is reached at $K_{\rm s}>20$ mag \citep{rubele15},
varying with local crowding levels. The stars we study in this work are
brighter than $K_{\rm s}=19.5$ mag, where completeness is always $>90$\%.

\begin{figure}
\includegraphics[width=\linewidth,clip=true]{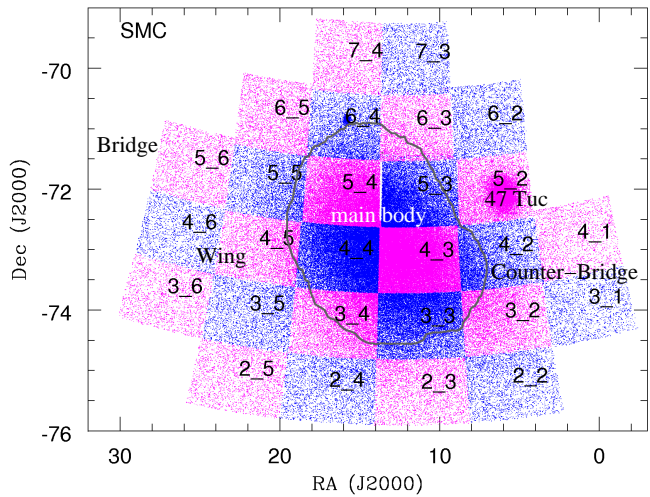}
\caption{Map of sources with $K_{\rm s}=16$--$18$ mag from the VMC survey area
around the SMC (the prominent dark area in the middle; the contour demarcates
the boundary between `inner' and `outer' regions). The VMC tiles are labelled.
The round overdensity seen at RA $=5\degr$, Dec $=-72\degr$ in tile SMC 5\_2
is caused by the Galactic globular cluster 47\,Tucan{\ae} (another Galactic
globular cluster, NGC\,362, is visible in tile SMC 6\_4 at RA $=16\degr$, Deg
$=-71\degr$). The gap between tiles SMC 5\_3 and SMC 5\_4 (seen at RA
$=13\rlap{.}{\degr}5$) is caused by the absence of data.}
\label{fig:str:smcall}
\end{figure}

The data used covers the entire VMC survey area for the SMC, encompassing 27
SMC tiles \citep[each covering almost uniformly an area of 1.5 deg$^2$ in
which each pixel received at least two exposures;][]{sutherland15}.
Figure~\ref{fig:str:smcall} shows a source density map of the SMC for the
$K_{\rm s}=16$--18 mag range. All coordinates quoted throughout this paper are
J2000. The gap between tiles SMC 5\_3 and SMC 5\_4 (RA $=13\rlap{.}{\degr}5$
Dec $=-72\degr$) is the result of the automatic centring of each tile in order
to include a sufficient number of guide stars (the gap is being covered with
new VISTA observations). One of the tiles could not be used in full (SMC 5\_2;
centred at RA $=6\rlap{.}{\degr}7$, Dec $=-71\rlap{.}{\degr}9$) because it
contains the Galactic globular cluster 47\,Tucan{\ae} causing significant
contamination of the region in the CMD where the RC is expected in the SMC
\citep{cioni16}. The excluded part of the tile is where RA
$<7\rlap{.}{\degr}8$ and Dec $<-71\rlap{.}{\degr}4$. Another Galactic globular
cluster, NGC\,362 is visible at RA $=16\degr$, Dec $=-71\degr$ but has
insignificant effect on our SMC RC analysis because of its larger distance vis
\`a vis 47\,Tucan{\ae}. The SMC has a very low stellar density outside of its
main body making the Galactic foreground more visible.

\begin{figure}
\includegraphics[width=\linewidth,clip=true]{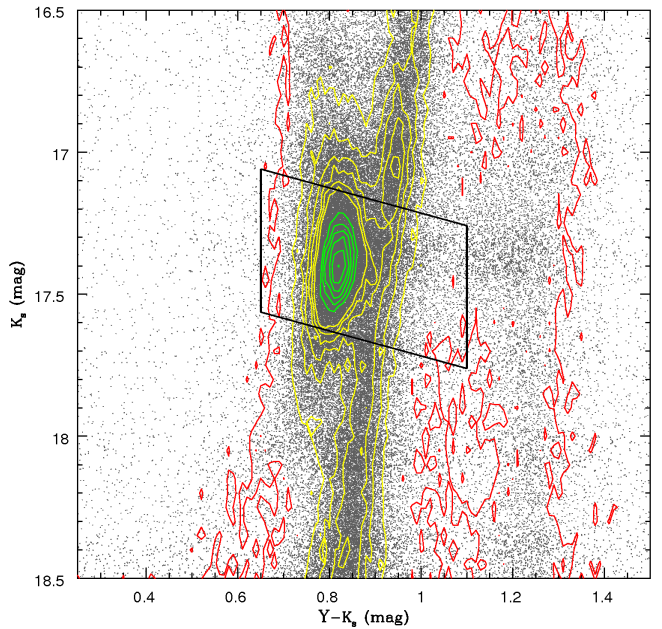}
\caption{$K_{\rm s}$ versus $Y-K_{\rm s}$ diagram around the RC in tile SMC 5\_3,
showing the paralellogram selection box inclined along the reddening vector
and limited at $(Y-K_{\rm s})=1.1$ mag to avoid Galactic foreground stars. The
density contours are at 1\% of peak (red), 5\%--30\% of peak with step 5\%
(yellow) and 50\%--100\% of peak with step 10\% (green).}
\label{fig:selection}
\end{figure}

We used point spread function (PSF) fitted photometry \citep[for full details
see][]{rubele12,rubele15} instead of the aperture photometry that is provided
by the VISTA Science Archive \citep[VSA;][]{cross12} as part of the VISTA
pipeline products. Apart from improved performance in crowded regions, the PSF
photometry was also found to be less susceptible to variations in seeing.

The tiles overlap by $\sim10$ per cent. As the photometric catalogues were
constructed on a tile-by-tile basis this means that some stars will have
multiple entries in the catalogue. This is not a problem, as our analysis is
not based on source counts. In fact it has the benefit that variations between
tiles will be reduced in these overlapping regions, and that any dispersion in
values introduced by multiple measurements is a measure of the accuracy with
which these measurements had been derived.

\section{Methods}\label{smc:meth}
\subsection{RC selection}\label{th:meth:cmd}

Each tile has its own RC star selection created using a two-dimensional
histogram (Hess diagram) method, described by \citet[Section 3.3]{tatton13}.
The only substantive difference with the procedures introduced by
\citet{tatton13} is that they made use of the $J$ band instead of the $Y$
band. The reason is that, at the time, problems were noticed with the $Y$ band
zero points that have meanwhile been resolved. With reliable $Y$-band
photometry now available, the $Y$ band is preferred over the $J$ band, in
conjunction with the $K_{\rm s}$ band, since the longer wavelength base helps
determine -- and correct for -- the amount of reddening by interstellar dust,
leading to more accurate reddening-corrected $K_{\rm s}$ values and thereby
more accurate distance modulus estimates. Both the $Y$-band and $J$-band data
are essentially complete for all RC stars; crowding and reddening lead to
negligible losses.

\begin{table*}
\caption{Tiles used and RC photometric properties. The columns list the tiles,
their centres, the number of sources with magnitude readings in all three
bands, the number of sources in the selected RC region followed by statistics
on these RC regions: contour peak position (bin size 0.05 and 0.01 mag for
$K_{\rm s}$ and $Y-K_{\rm s}$ respectively) and the $Y-K_{\rm s}$ mean, standard
deviation (SD) and median values.}
\begin{tabular}{cccrrccccc}
\hline\hline
\noalign{\smallskip}
Tile & RA$_{\rm J2000}$ ($\degr$) & Dec$_{\rm J2000}$ ($\degr$) & \multicolumn{2}{c}{Sources} & \multicolumn{2}{c}{--- Contour peak ---} & Mean & SD & Median \\
 & \degr & \degr & Total & RC & $K_{\rm s}$ (mag) & \multicolumn{4}{c}{\llap{--}------------ $Y-K_{\rm s}$ (mag) --------------} \\
\noalign{\smallskip}
\hline
\noalign{\smallskip}
SMC 2\_2 & 05.4330 & $-$75.2012 &   84627 &   5486 & 17.45 & 0.82 & 0.822 & 0.065 & 0.810 \\
SMC 2\_3 & 11.1496 & $-$75.3037 &  110576 &  10548 & 17.45 & 0.78 & 0.816 & 0.060 & 0.807 \\
SMC 2\_4 & 16.8911 & $-$75.2666 &  137434 &  10941 & 17.50 & 0.80 & 0.906 & 0.179 & 0.830 \\
SMC 2\_5 & 22.5526 & $-$75.0910 &   72423 &   5749 & 17.40 & 0.80 & 0.839 & 0.076 & 0.823 \\
SMC 3\_1 & 00.6663 & $-$73.8922 &   76079 &   2068 & 17.40 & 0.75 & 0.792 & 0.075 & 0.780 \\
SMC 3\_2 & 05.8981 & $-$74.1159 &  190401 &  18013 & 17.40 & 0.77 & 0.793 & 0.059 & 0.782 \\
SMC 3\_3 & 11.2329 & $-$74.2117 & 1631078 &  56702 & 17.40 & 0.82 & 0.836 & 0.059 & 0.830 \\
SMC 3\_4 & 16.5880 & $-$74.1774 &  288092 &  29794 & 17.40 & 0.83 & 0.840 & 0.058 & 0.834 \\
SMC 3\_5 & 21.8784 & $-$74.0137 &  128290 &   7314 & 17.40 & 0.83 & 0.845 & 0.072 & 0.839 \\
SMC 3\_6 & 27.0255 & $-$73.7245 &   97367 &   3542 & 17.40 & 0.78 & 0.818 & 0.074 & 0.803 \\
SMC 4\_1 & 01.3911 & $-$72.8200 &   75399 &   1895 & 17.50 & 0.76 & 0.779 & 0.083 & 0.758 \\
SMC 4\_2 & 06.3087 & $-$73.0299 &  245652 &  21829 & 17.45 & 0.78 & 0.809 & 0.070 & 0.794 \\
SMC 4\_3 & 11.3112 & $-$73.1198 &  845029 & 125704 & 17.40 & 0.86 & 0.884 & 0.075 & 0.878 \\
SMC 4\_4 & 16.3303 & $-$73.0876 &  693135 &  82330 & 17.40 & 0.86 & 0.872 & 0.075 & 0.868 \\
SMC 4\_5 & 21.2959 & $-$72.9339 &  201070 &  11533 & 17.40 & 0.84 & 0.848 & 0.066 & 0.840 \\
SMC 4\_6 & 26.1438 & $-$72.6624 &   70533 &   4310 & 17.00 & 0.81 & 0.828 & 0.069 & 0.815 \\
SMC 5\_2 & 06.6737 & $-$71.9433 &   42860 &   3449 & 17.50 & 0.78 & 0.807 & 0.070 & 0.775 \\
SMC 5\_3 & 11.2043 & $-$72.0267 & 2174210 &  45536 & 17.40 & 0.82 & 0.840 & 0.064 & 0.829 \\
SMC 5\_4 & 16.1088 & $-$71.9975 &  536306 &  59231 & 17.35 & 0.84 & 0.858 & 0.069 & 0.849 \\
SMC 5\_5 & 20.7706 & $-$71.8633 &  185805 &  14004 & 17.25 & 0.82 & 0.841 & 0.067 & 0.828 \\
SMC 5\_6 & 25.3700 & $-$71.5964 &   93924 &   4151 & 17.00 & 0.77 & 0.799 & 0.068 & 0.782 \\
SMC 6\_2 & 06.9165 & $-$70.8535 &   91744 &   3822 & 17.50 & 0.77 & 0.811 & 0.073 & 0.775 \\
SMC 6\_3 & 11.4532 & $-$70.9356 &  123840 &   8814 & 17.50 & 0.79 & 0.830 & 0.061 & 0.817 \\
SMC 6\_4 & 15.9581 & $-$70.8929 &  155280 &  12598 & 17.30 & 0.79 & 0.811 & 0.066 & 0.798 \\
SMC 6\_5 & 20.3437 & $-$70.7697 &  118002 &   7433 & 17.00 & 0.77 & 0.796 & 0.066 & 0.781 \\
SMC 7\_3 & 11.5197 & $-$69.8439 &   67270 &   3428 & 17.55 & 0.75 & 0.783 & 0.067 & 0.767 \\
SMC 7\_4 & 15.7520 & $-$69.8162 &   77063 &   3632 & 17.50 & 0.78 & 0.798 & 0.066 & 0.783 \\
\noalign{\smallskip}
\hline
\end{tabular}
\label{tab:smclist}
\end{table*}

To summarise, each tile has a RC selection box with a centre and size
determined by the peak and width of the distribution of stars in the CMD and
with a colour gradient determined by the reddening vector
(Fig.~\ref{fig:selection}). Contour plots are used to determine the centre and
magnitude range within the box whereas histograms inside this box are used to
determine the colour range. These values are summarised in
Table~\ref{tab:smclist}. A ``red'' boundary to the selection box was imposed
in order to avoid running into the vertical sequence of foreground stars
(Fig.~\ref{fig:selection}); owing to the metal-poor SMC stars being relatively
blue, this was possible whereas in the LMC the two populations overlap
\citep[see Figure C.1 in][]{tatton13}. The foreground population appears to
have an excess of stars only slightly brighter than the RC stars in
Fig.~\ref{fig:selection}. It is not clear what is the cause of this (they
cannot be blended or reddened RC stars), but it does not affect our analysis.

In total 561,843 RC star candidates were found. The `inner' and `outer' SMC
regions are defined by a density contour map of the SMC for the RC stars,
where the outer SMC lies outside of the 20\% contour level -- this contains
161,100 sources.

\begin{figure*}
\includegraphics[angle=0, width=\linewidth]{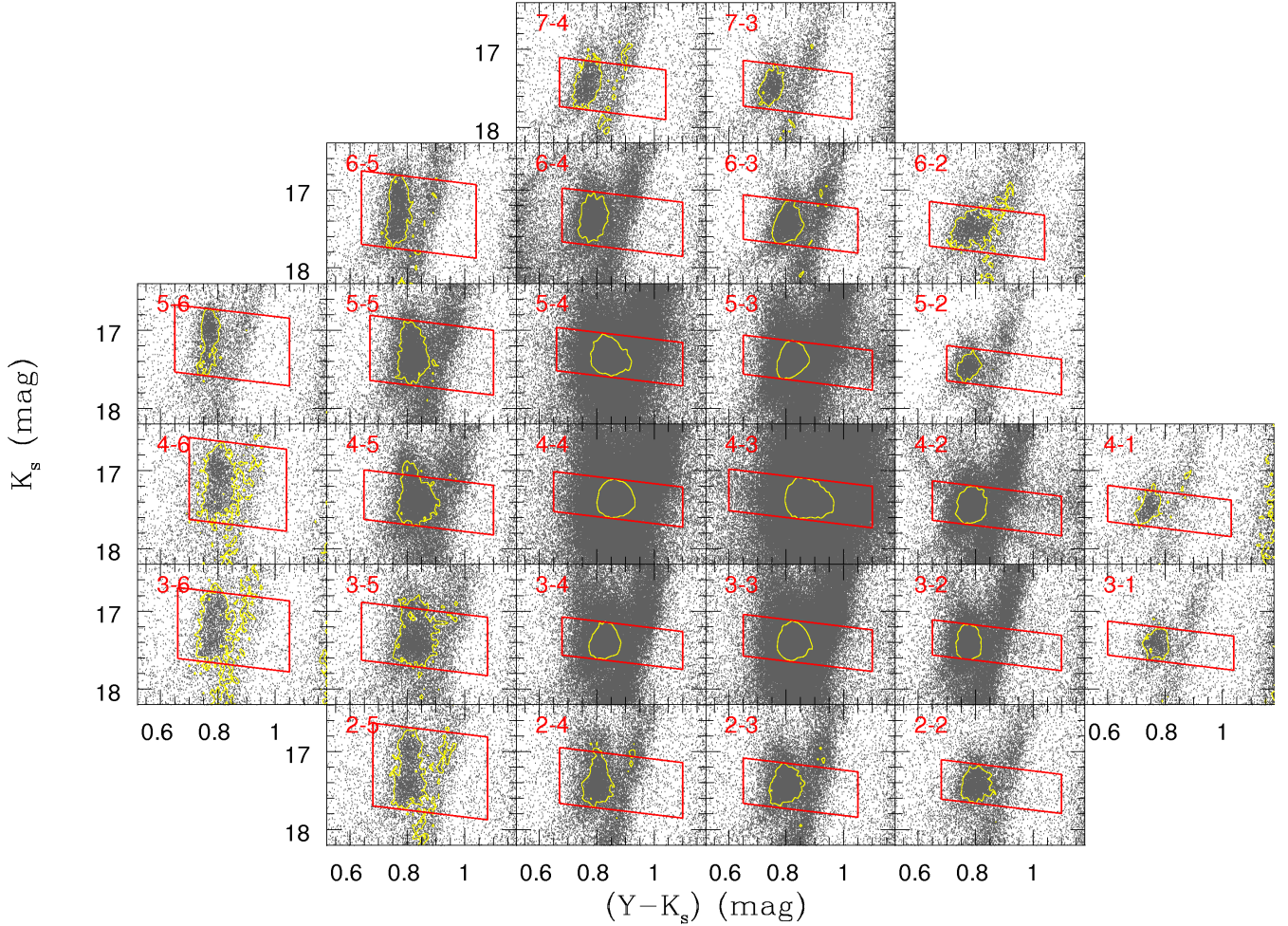}
\caption{CMDs for all tiles, each with their RC selection box and contour at
30\% of the RC peak within the CMD.}
\label{fig:smc:selecta}
\end{figure*}

\subsection{Population effects}\label{smc:pop}

There are two possible population effects that could affect the assumptions
made in the use of individual RC stars: intrinsic colour and brightness
variations (or spread) resulting from age and/or metallicity variations (or
spread), and contamination from stars that are not RC stars.

To start with the former, it is true that even in the K-band the RC brightness
has the potential to vary by more than 0.5 mag, becoming brighter with younger
age and higher metallicity \citep[e.g., Figures 1 and 3 in][]{salaris02}. This
could lead to under-estimation of the distance modulus (or introduce an
apparent foreground population) or -- in the case of a mix of populations of a
range in age and metallicity -- an increased apparent depth. Because these
population variations are largely due to bolometric correction variations the
intrinsic colours will also change, which could lead to errors in the
reddening determination and correction.

However, in reality the situation is not as extreme as one might envisage,
because the SMC has been forming stars over many Gyr, gradually building up
its metallicity. In combination with the mass-dependent stellar birth
frequency and evolutionary timescales this results in a dominant age and
metallicity corresponding to the assumed RC. Figure 13 in \citet{rubele18}
proves this point: based on star formation and chemical evolution histories
derived from the CMDs across the SMC, the RC stars effectively comprise
intermediate age and only a narrow range in (typical SMC) metallicity,
suggesting there is no cause for concern. In particular, there is no evidence
for a significant contribution of (relatively) young RC stars. The RC does
have a finite spread, even for a simple stellar population, but the core of
the RC is very sharp ($\Delta K_{\rm s}\ll 0.1$ mag; Figure 6 in
\citet{rubele18}) and comparable to the photometric spread. The spread derived
from the inner two quartiles of the brightness distribution will measure
mainly how much the core of the RC is spread, and be insensitive to intrinsic
tails of the RC brightness distribution; while maps of the spatial
distribution in the outer quartiles will show if and where there is an excess
of stars that are brighter/fainter because they are nearer/farther.

As for the latter effect, contamination can arise principally from RGB stars
and foreground stars. Figure~\ref{fig:smc:selecta} shows evidence for this,
and how it may vary across the SMC. The contaminating foreground population
comprises relatively cool, metal-rich stars in the Milky Way disc, seen along
the red edge of the CMDs around $(Y-K_{\rm s})\approx1.4$ mag. Here we are
helped by the fact that SMC stars, even the RC stars, are relatively warm and
metal-poor. This separates the two populations in colour, and given that the
reddening is also relatively mild in the not very dusty metal-poor ISM of the
SMC the two do not overlap. Nevertheless, the RC selection boxes are truncated
at the red side so as to avoid the foreground, sacrificing the possible
detection of rare highly-reddened RC stars -- even in crowded fields with
relatively dense, possibly dusty ISM such as SMC 4\_3 there is no indication
that such stars exist in abundance.

As is clear in Figure~\ref{fig:smc:selecta}, however, the RGB within the SMC
is crossed by the selection box. While the bulk of the RC is offset to bluer
colours with respect to the bulk of the RGB, we had decided not to restrict
the selection boxes further so as to avoid the RGB altogether, as this would
introduce a stronger bias against reddened RC stars that could preferentially
be at somewhat larger distances. Consequently, there is a fraction of the RC
sample that are not in actual fact RC stars but RGB stars. The question is how
this affects our analysis.

Firstly, as is clear from Figure~\ref{fig:smc:selecta}, the RC is much more
densely populated within the CMD than the RGB at similar brightness. The
contours delineate 30\% of the RC peak, and the RGB CMD density is well below
that. Secondly, the RGB is populated much more uniformly in brightness, with a
shallow and smooth gradient, in comparison to the RC (``branch'' versus
``clump''). This means that the RGB contamination should not lead to large
deviations in the derived median distance modulus (as used in
Section~\ref{smc:plane}). It could have a larger effect on the apparent depth,
but we mitigate against this by considering the brightness distribution of the
RC stars excluding the tails of the distributions (see Section~\ref{smc:res}).

The largest effect could be on the derived RC reddening distributions. There
is little reason to expect the spatial distribution in reddening (discussed in
Section~\ref{smc:red}) to be affected substantially as the ratio of RC to RGB
stars varies little across the SMC. However, we must look out for possible
increases in stars that appear to be reddened significantly but that are
actually (relatively) unreddened RGB stars. We discuss evidence for this, and
quantify its effect, in Section~\ref{smc:dust}.

\subsection{Dereddening}\label{smc:dered}

It is possible to obtain a value for the $K_{\rm s}$ magnitude of an individual
RC star free from extinction, based on its $Y-K_{\rm s}$ colour excess,
$E(Y-K_{\rm s})$, with respect to its intrinsic colour, $(Y-K_{\rm s})_0$. A
value of $(Y-K_{\rm s})_0=0.7625$ mag was determined from the average values
using isochrones \citep[][converted to the VISTA photometric system by
\citealt{rubele12}]{marigo08,girardi10} that represent the SFH derived for the
SMC from VMC PSF photometry by \citet{rubele15}. In fact, the photometric
properties of the bulk of the RC stars are not significantly affected by the
effects of age or metallicity, especially in the near-IR
\citep{alves00,onozato19}. This fiducial colour agrees very well with the
observed blue end of the range of median values for the RC peak
($(Y-K_{\rm s})\approx 0.78$ mag; see Table~\ref{tab:smclist}); redder values
indicate the effects of reddening.

The de-reddened magnitude can be expressed as:
\begin{equation}
(K_{\rm s})_0=K_{\rm s}-G\times\,E(Y-K_{\rm s}),
\end{equation}
where $(K_{\rm s})_0$ is the resulting de-reddened magnitude and $G$ is the
ratio of total to selective extinction (the `reddening vector'):
\begin{equation}
G=\frac{A_{K{\rm s}}/A_V}{A_Y/A_V-A_{K{\rm s}}/A_V}.
\end{equation}
Using the $A_\lambda/A_V$ values for the VISTA filters \citep[][their Table
2 based on the \citet{cardelli89} extinction law]{tatton13} we obtain
$G=0.443$ and $A_V=3.69\times E(Y-K_{\rm s})$.

De-reddening is not carried out if $E(Y-K_{\rm s})<0$, which would have
resulted from photometric scatter and intrinsic spread and not reddening. We
do include these negative colour excess values in the reddening map so as not
to bias it. We do also include those RC stars in the 3D structure analysis,
without changes to their colours and magnitudes; a similar number of stars
would be expected to the red of the RC fiducial colour, but all of those are
treated as being reddened and their colours and magnitudes are corrected. The
latter will introduce an offset in the distances estimated for these stars
(bringing them nearer to us), but this is a small effect
($\Delta E(Y-K_{\rm s})\sim0.03$ mag $\equiv 0.013$ mag), diluted by at least
as many sources not affected, and systematic across the entire survey area.

\section{Reddening map}\label{smc:red}

\begin{figure*}
\includegraphics[angle=0,clip=true,width=0.51\linewidth]{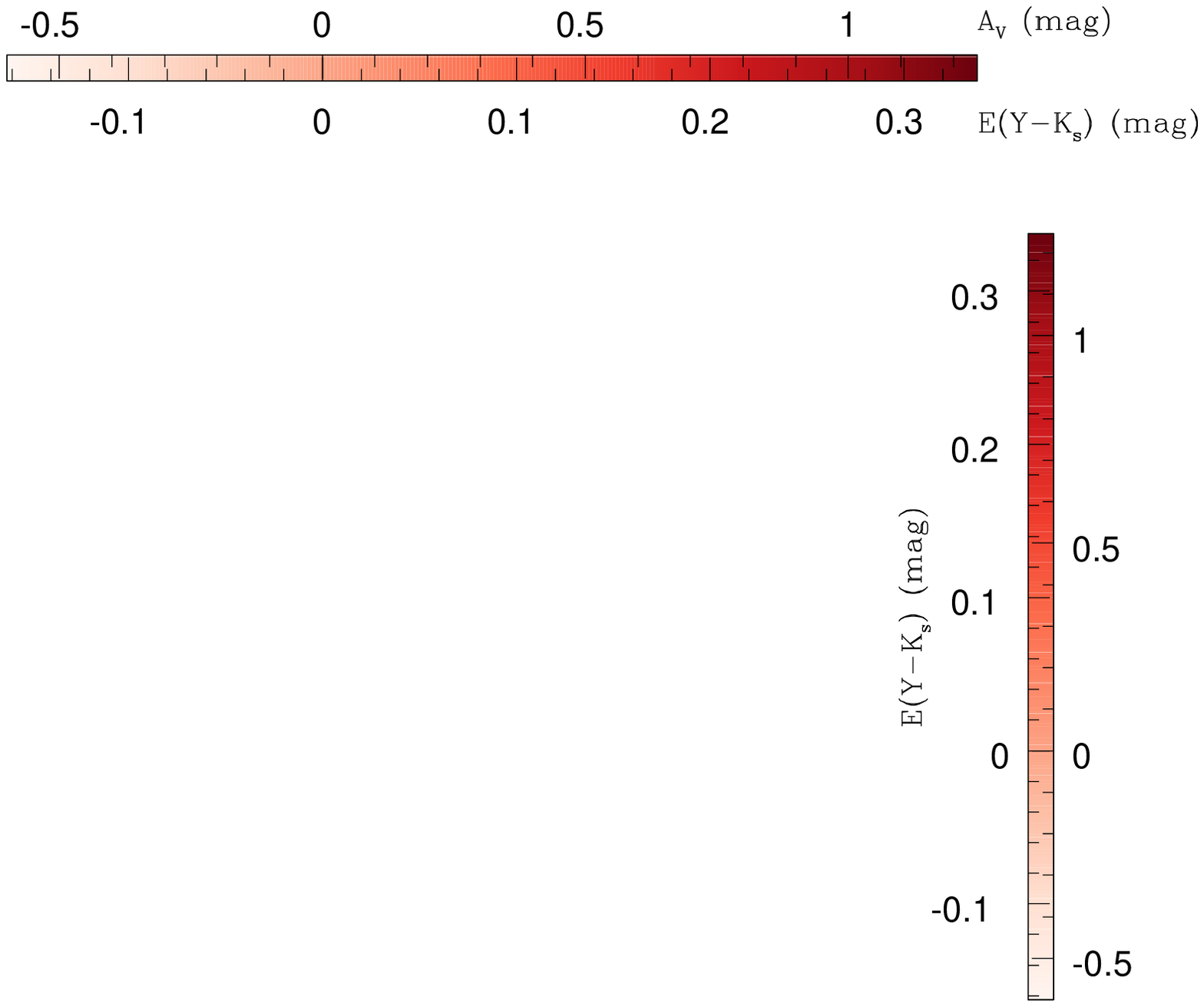}
\includegraphics[angle=0,width=0.487\linewidth]{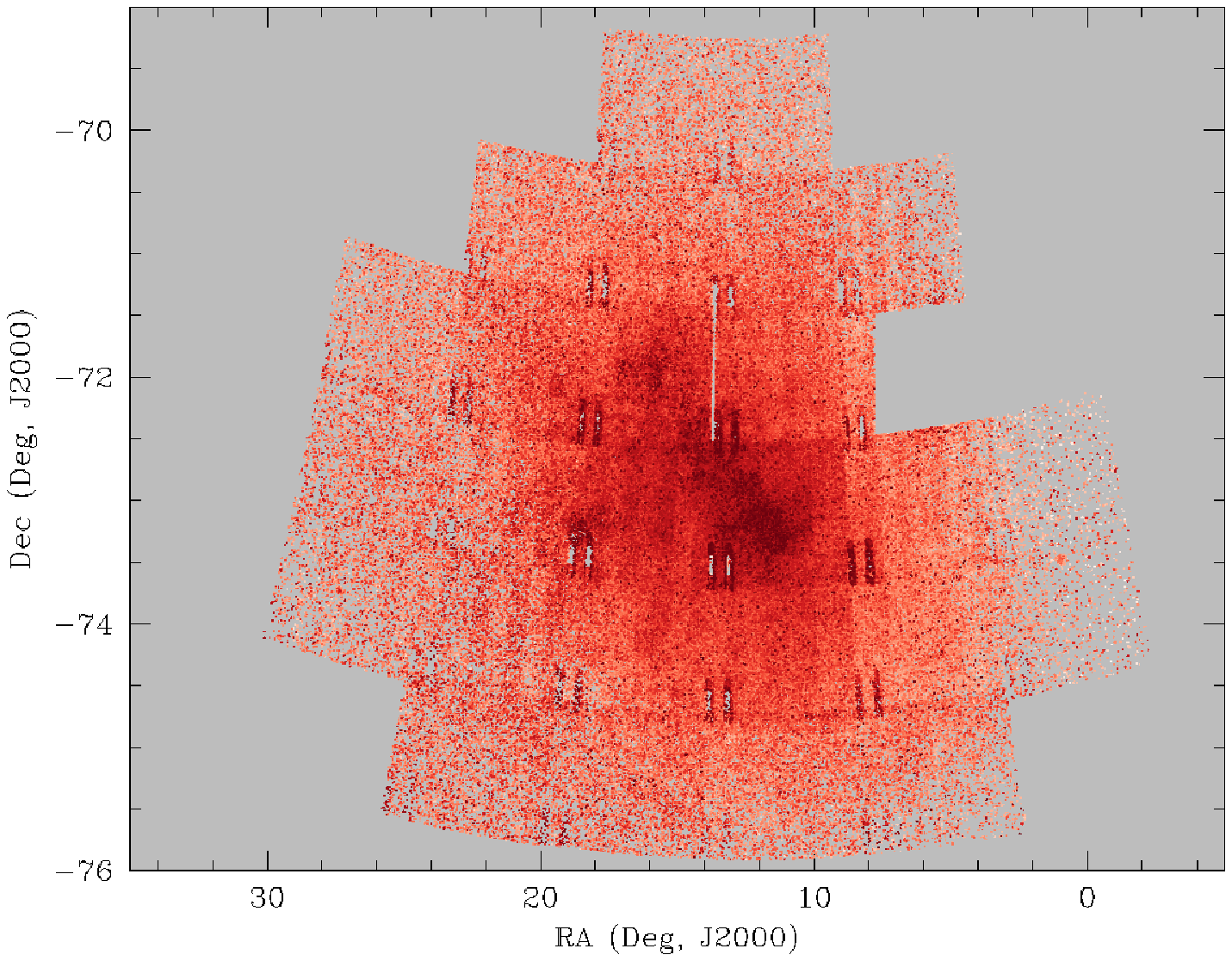}
\includegraphics[angle=0,clip=true,width=0.5\linewidth]{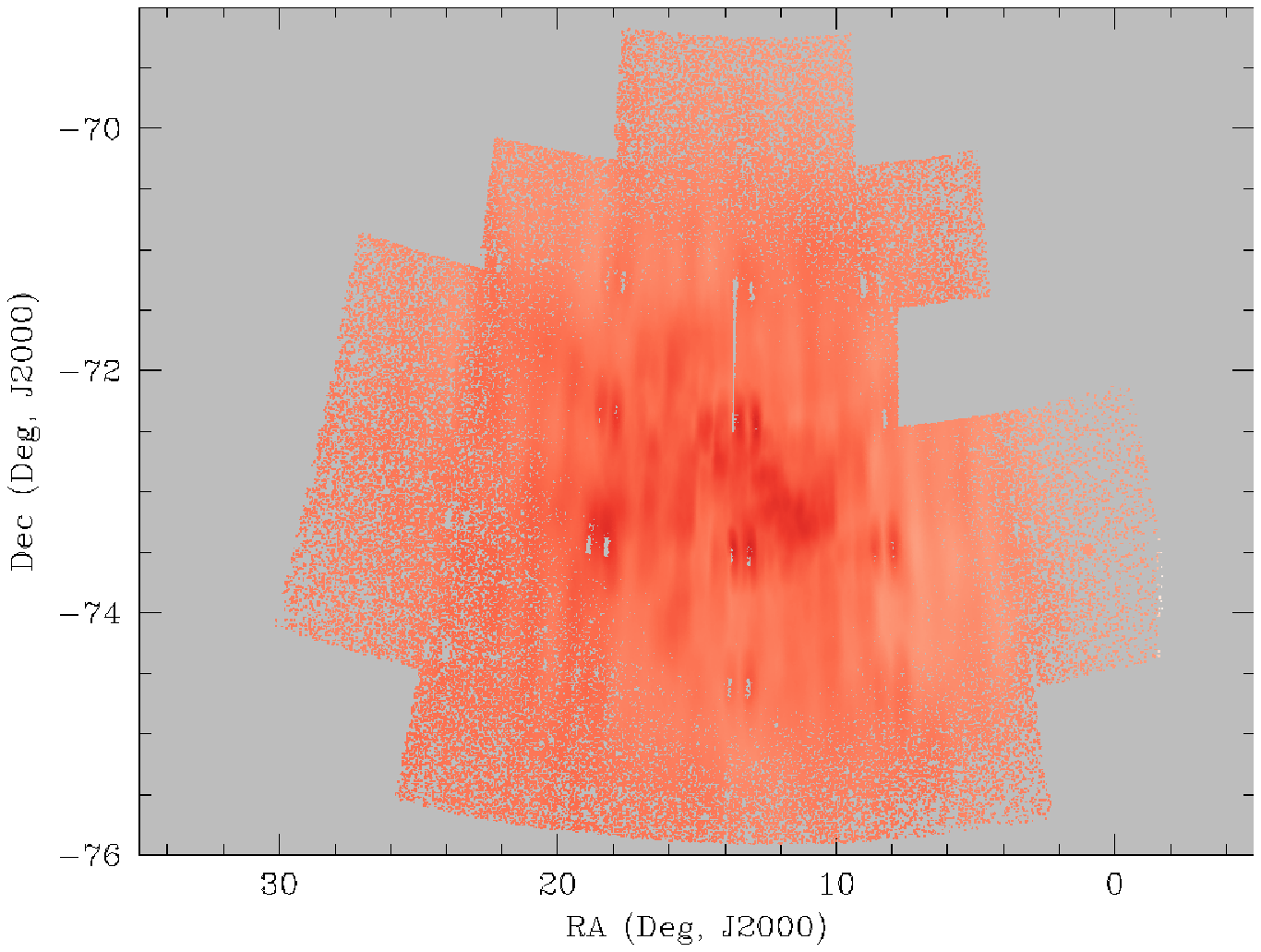}
\caption{Reddening map of the SMC in $E(Y-K_{\rm s})$ (top side of legend shows
this converted into $A_V$ magnitude) from RC stars. The vertical white strip
around RA $=14\degr$ and Dec $=-72\degr$ lacks data, while the repeated
pattern of pairs of short bars are caused by detector 16. The right panel
shows the mean reddening values for the nearest 1000 RC stars.}
\label{fig:smc:red}
\end{figure*}

Figure~\ref{fig:smc:red} shows the reddening map of the entire SMC on the
$E(Y-K_{\rm s})$ scale along with the conversion to $A_V$, both on a
star-by-star basis (left panel) and as mean values for the nearest 1000 RC
stars (right panel). The latter results in a varying angular resolution, from
$\sim10^\prime$ in the densest parts of the SMC (175 pc at a distance of 60 kpc
corresponding to the bulk of the SMC ) to $\sim1^\circ$ in the outskirts (1.75
pc at a distance of 100 pc for a typical Galactic dust layer).
\citet{gonzalez12} determined that 200 is an appropriate minimum number of RC
stars over which to average based on the Gaussian distributions over
$(J-K_{\rm s})$ colour of RC stars seen in the direction of the Galactic Bulge.
The larger number of stars (1000) chosen here reflects the relatively low
reddening in the direction of the SMC and larger distance of RC stars within
the SMC, thus increasing the statistical errors and the stochastic width of
the Gaussian distributions which, in turn, necessitates a larger number of
sightlines to be combined. Both reddening maps are made available at the
Centre de Donn\'ees astronomiques de Strasbourg (CDS).

The field edges stand out more clearly than in Figure~\ref{fig:str:smcall} as
a result of the varying RC selection box (see Table~\ref{tab:smclist}). The
streaks of missing data and occasionally high extinction in a North--South
direction are due to the variable quantum efficiency of VISTA detector 16
\footnote{
http://casu.ast.cam.ac.uk/surveys-projects/vista/technical/known-issues}. This
is especially prevalent in tiles SMC 3\_2 (centred at RA $=6\degr$, Dec
$=-74\degr$) and SMC 5\_4 (centred at RA $=16\degr$, Dec $=-72\degr$).

High extinction is mainly seen along the dense main body of the SMC, though it
becomes more patchy towards north--western parts covered largely by tile SMC
5\_3 (RA $\simeq12\degr$, Dec $\simeq-72\degr$). The strongest features
however are seen almost exclusively within tile SMC 4\_3 (RA
$=11\rlap{.}{\degr}3$, Dec $=-73\rlap{.}{\degr}1$). Evidently, the dust is
concentrated in the main body where star formation is also most intense
\citep[e.g.,][]{bolatto11}. In general for the whole SMC, lower extinction is
seen in the Wing to the East of the main body.

\section{Three-dimensional distribution of RC stars}\label{smc:3d}
\subsection{Regional census}\label{smc:cen}

\begin{figure*}
\includegraphics[angle=0,clip=true, width=0.49\linewidth]{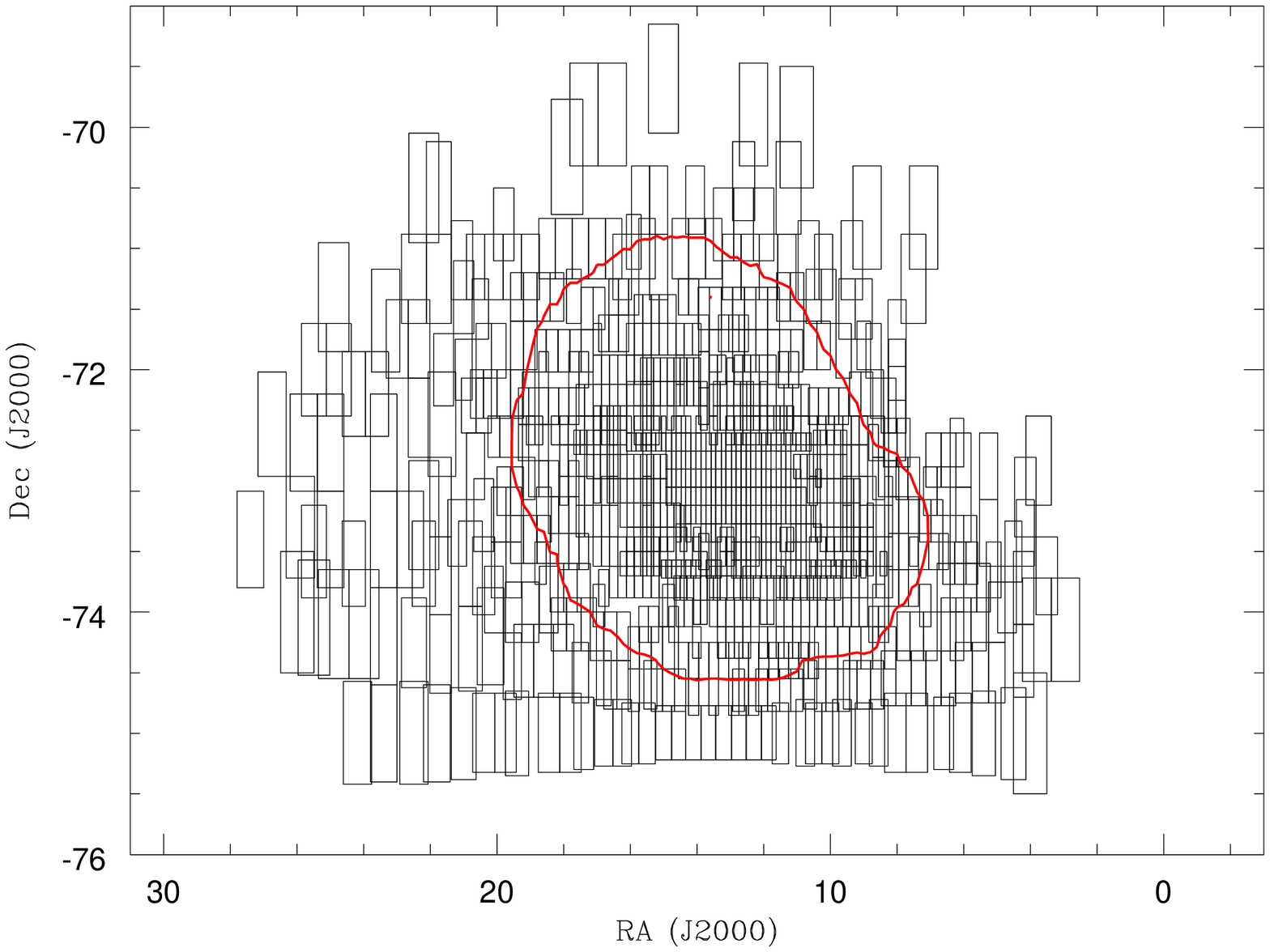}
\includegraphics[angle=0,clip=true, width=0.49\linewidth]{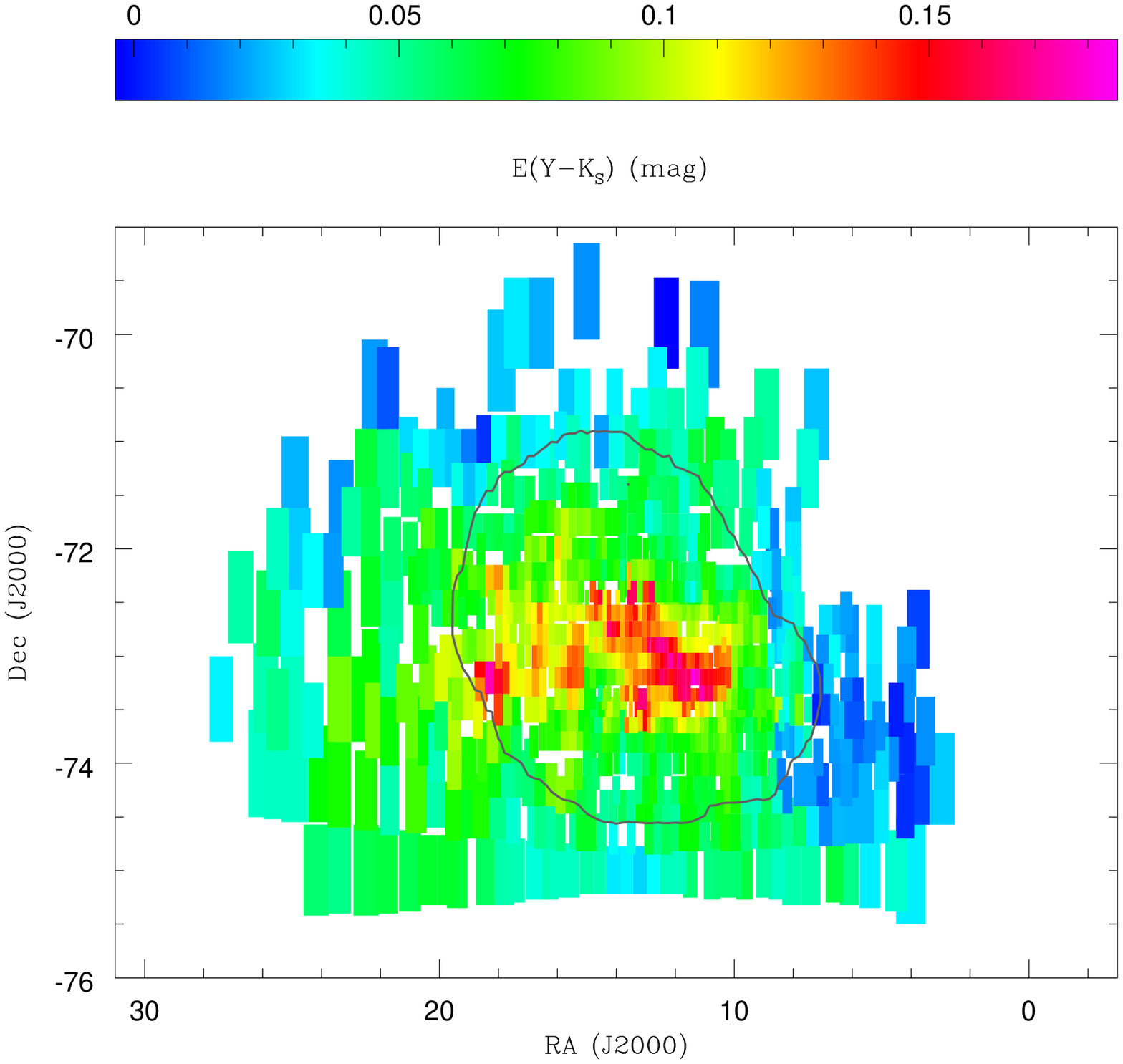}
\includegraphics[angle=0,clip=true, width=0.49\linewidth]{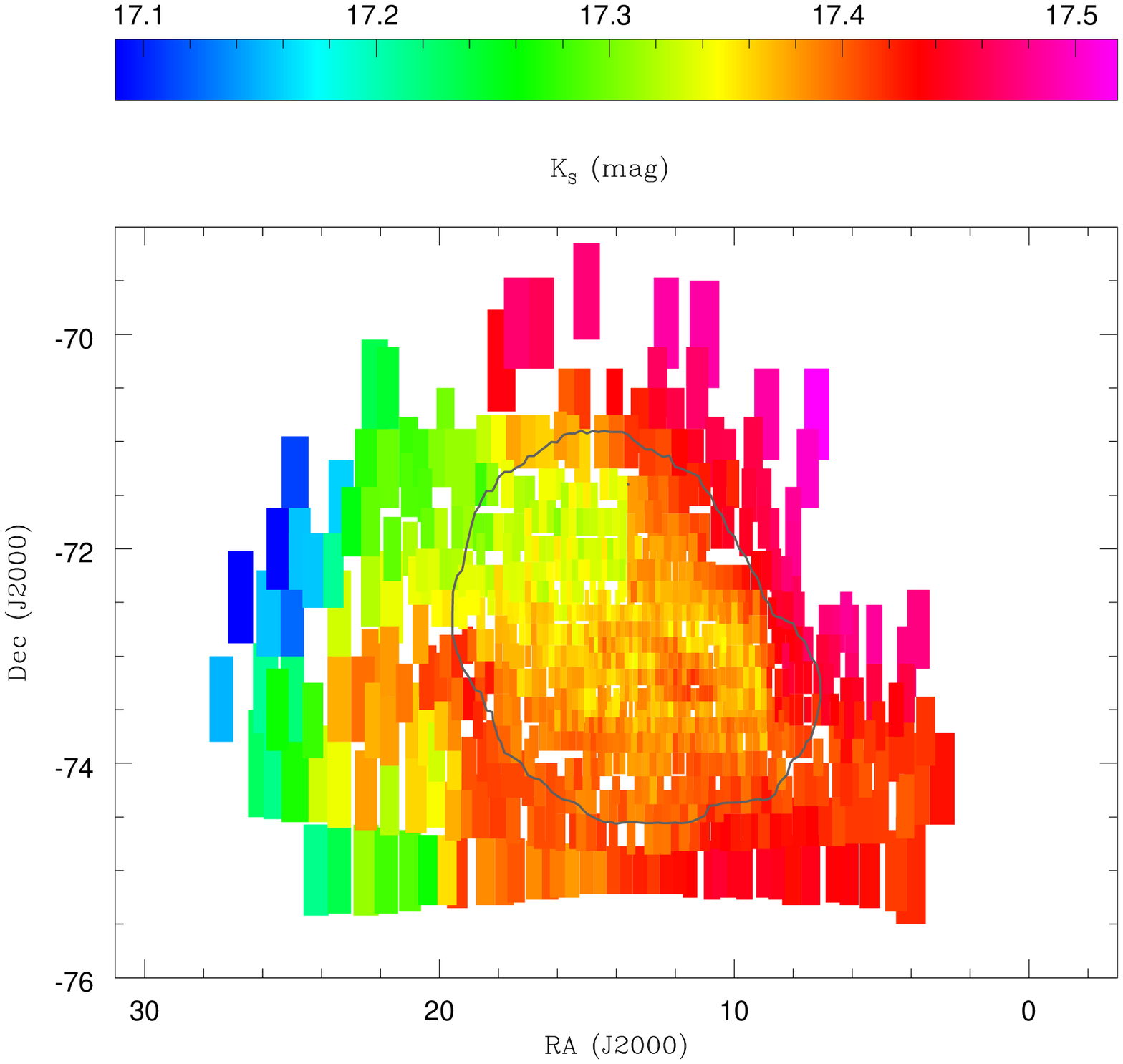}
\includegraphics[angle=0,clip=true, width=0.49\linewidth]{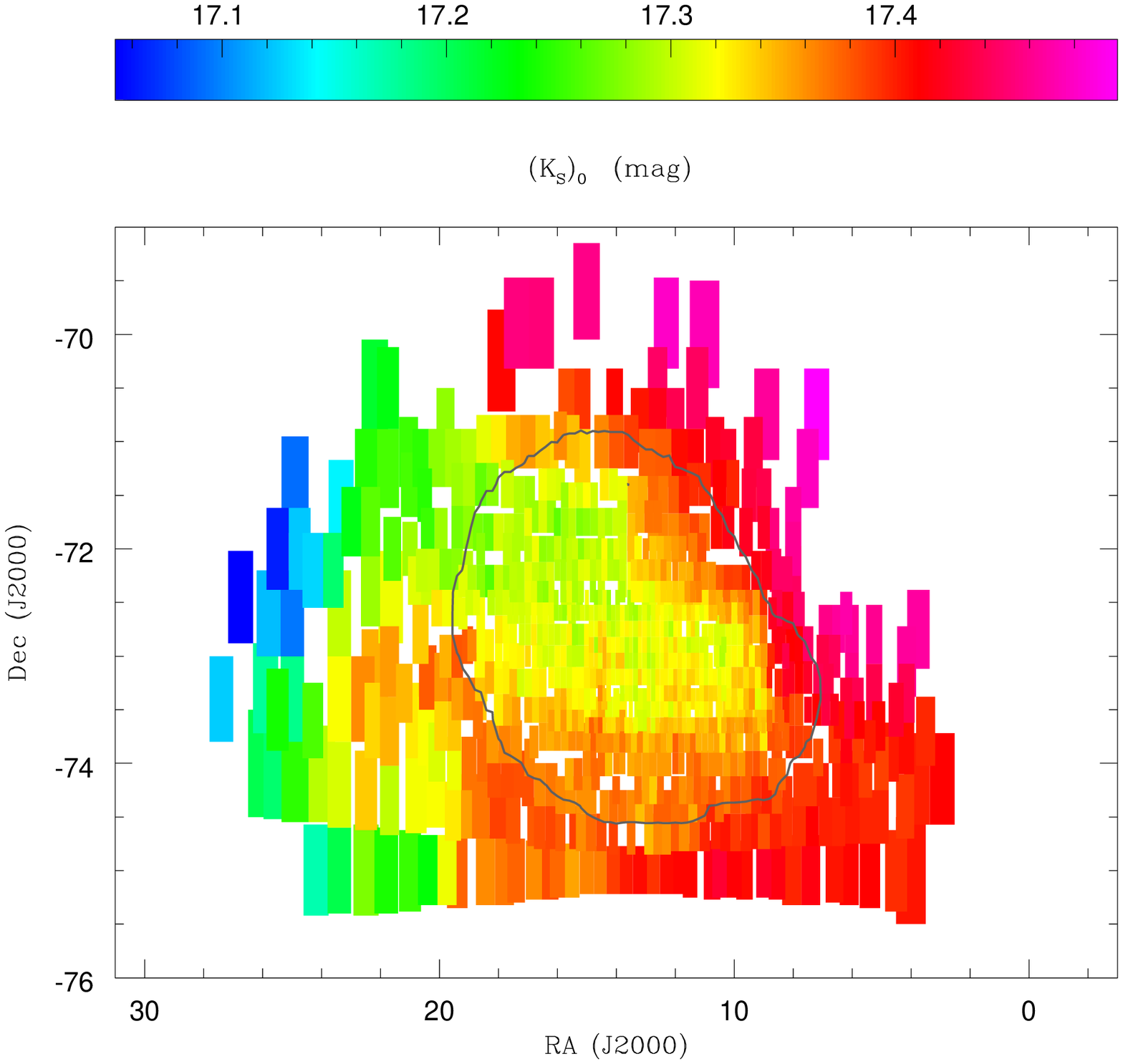}
\caption{SMC maps of RC stars based on the regional census method. Shown are
the outlines of the sub-regions (top--left), median $Y-K_{\rm s}$ colour
(top--right), median $K_{\rm s}$ magnitudes before (bottom--left) and after
(bottom--right) de-reddening. The contour represents the demarcation line
between the inner and outer SMC fields.}
\label{fig:smc:cen}
\end{figure*}

As a preliminary exploration of structure, the SMC RC star sample is divided
into binned regions whose sizes vary with the stellar density. Owing to some
SMC regions having very low stellar density this results in an incomplete
coverage of the SMC, particularly within the Wing. Each cell in this regional
census contains 500--2000 sources and has a size of 0.15--1 deg$^2$. This
produces a total of 976 sub-regions, shown in Figure~\ref{fig:smc:cen}.

As expected, the median $(Y-K_{\rm s})$ colour follows the general pattern of
the reddening map, where the inner SMC is redder than the outer regions. The
representation in this figure, however, places more emphasis on high
extinction regions -- notably the area around RA $=18\rlap{.}{\degr}5$, Dec
$=-73\rlap{.}{\degr}2$ associated with the N\,83/84 star-formation complex.

From the de-reddened magnitudes (Figure~\ref{fig:smc:cen}, bottom right) it
can be seen that the eastern Wing lies closer to us than the main body of the
SMC, and the western periphery of the SMC is the furthest away. This appears
to confirm the findings of \citet{bica15}, who measured shorter distances to
clusters within the Bridge (40--48 kpc). But the RC magnitude trend we see
here is not a straightforward East--West gradient.

\subsection{SMC line-of-sight depth}\label{smc:los}

There is a strong motivation to examine the line-of-sight depth in the SMC
because it has been shown to have a thicker north--eastern side
\citep[e.g.,][]{kapakos12,nidever13a,subramanian17}.

Assuming that an RC star with $K_{\rm s}=17.304$ mag corresponds to a distance
of 61 kpc (determined in the same way as the intrinsic colour in
Section~\ref{smc:dered}), we used the dispersion of distances in a binned
region as a proxy for its thickness -- i.e.\ line-of-sight depth.

\begin{figure}
\includegraphics[width=\linewidth,clip=true]{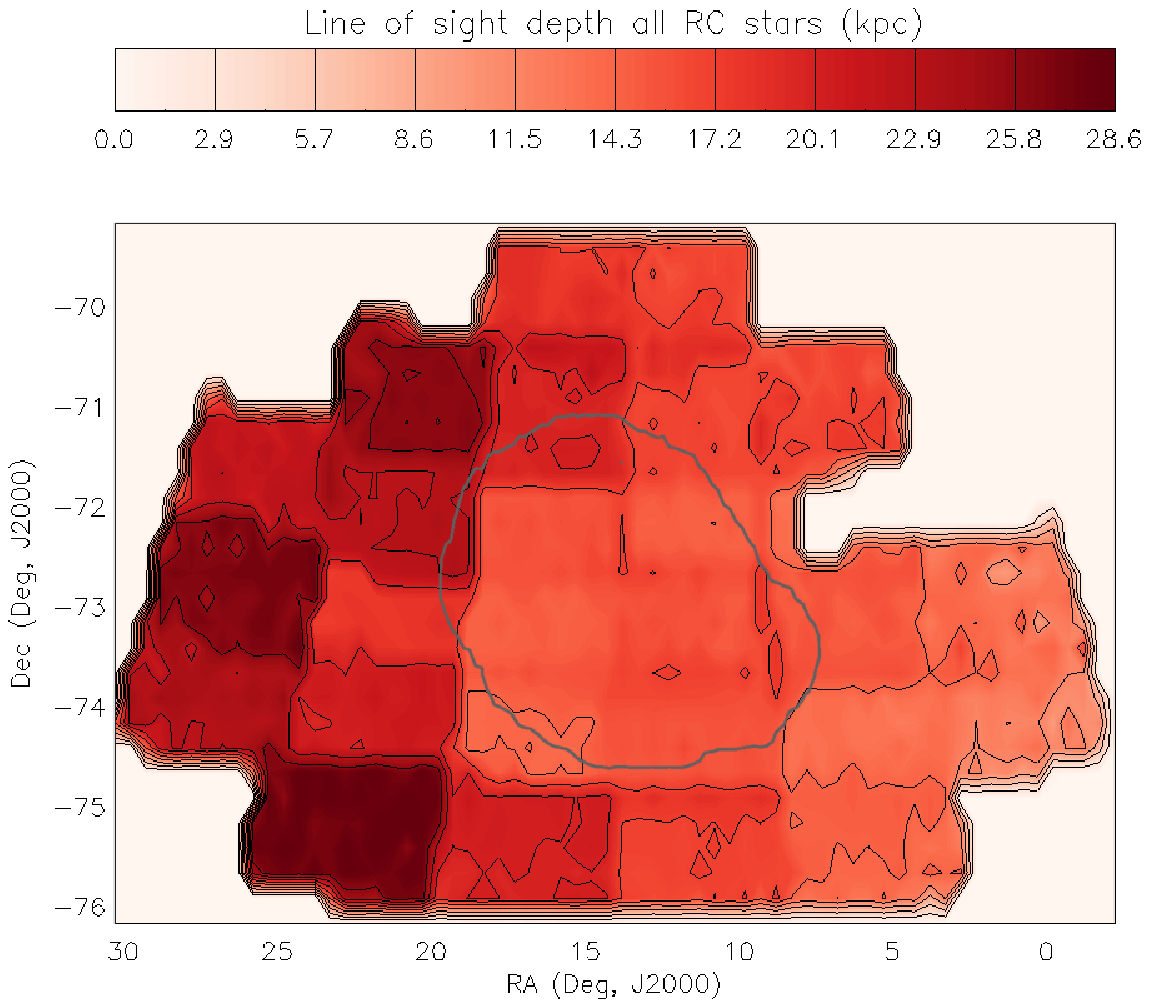}\\
\includegraphics[width=\linewidth,clip=true]{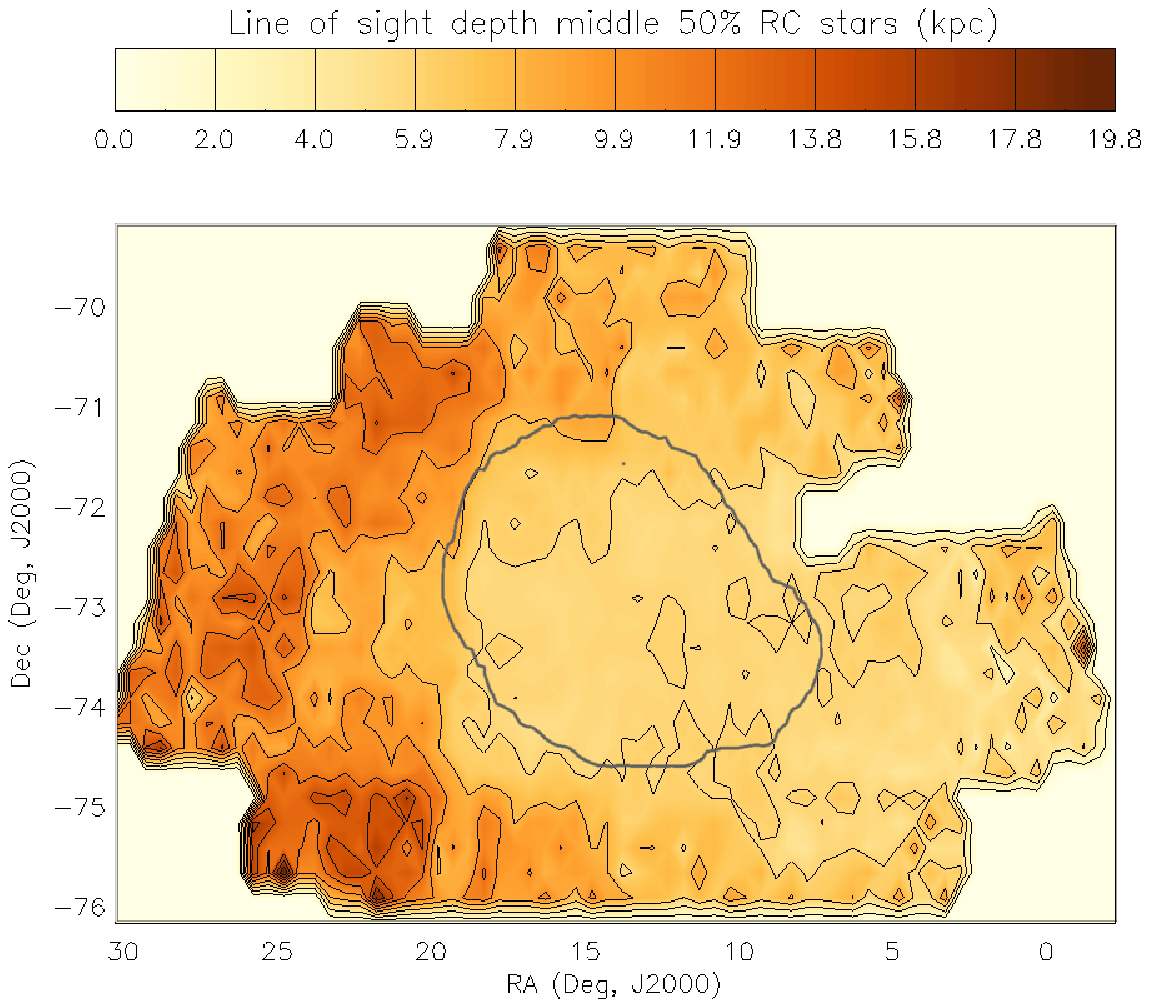}
\caption{Line-of-sight depth of the SMC (in kpc) from RC stars, for all
sources (top) and central quartiles (bottom). The bin size is
$0\rlap{.}{\degr}5$ in RA and $0\rlap{.}{\degr}25$ in Dec.}
\label{fig:str:smcdepth}
\end{figure}

For binned regions of $0\rlap{.}{\degr}5$ in RA and $0\rlap{.}{\degr}25$ in
Dec, we look at this in two ways: first, the overall thickness range covered
by all RC stars and, second, the thickness range of the two central quartiles
of the distance distribution. While the former may over-estimate the
line-of-sight depth due to contamination of the luminosity distribution by a
`continuum' of RGB stars, the latter likely under-estimates it since the RC
luminosity distribution will have become truncated. Contour maps of these
line-of-sight depth measures are shown in Figure~\ref{fig:str:smcdepth}.

The depth values themselves are typical of what is found in the literature,
with central depths of $\sim6$--14 kpc \citep{crowl01,subramanian12} and
eastern depths of up to $\sim23$ kpc towards the Bridge \citep{nidever13a}.

When considering all sources one can see the effect of RC star selection and
de-reddening having been conducted on a tile-by-tile basis, with the contour
levels typically changing between tile edges. There is still significant
line-of-sight depth exhibited in the central-quartiles sample. The major
contributors to this depth are mostly the Wing regions in the North--East and
South--East, but this is not a continuous band.

\begin{figure}
\includegraphics[width=\linewidth,clip=true]{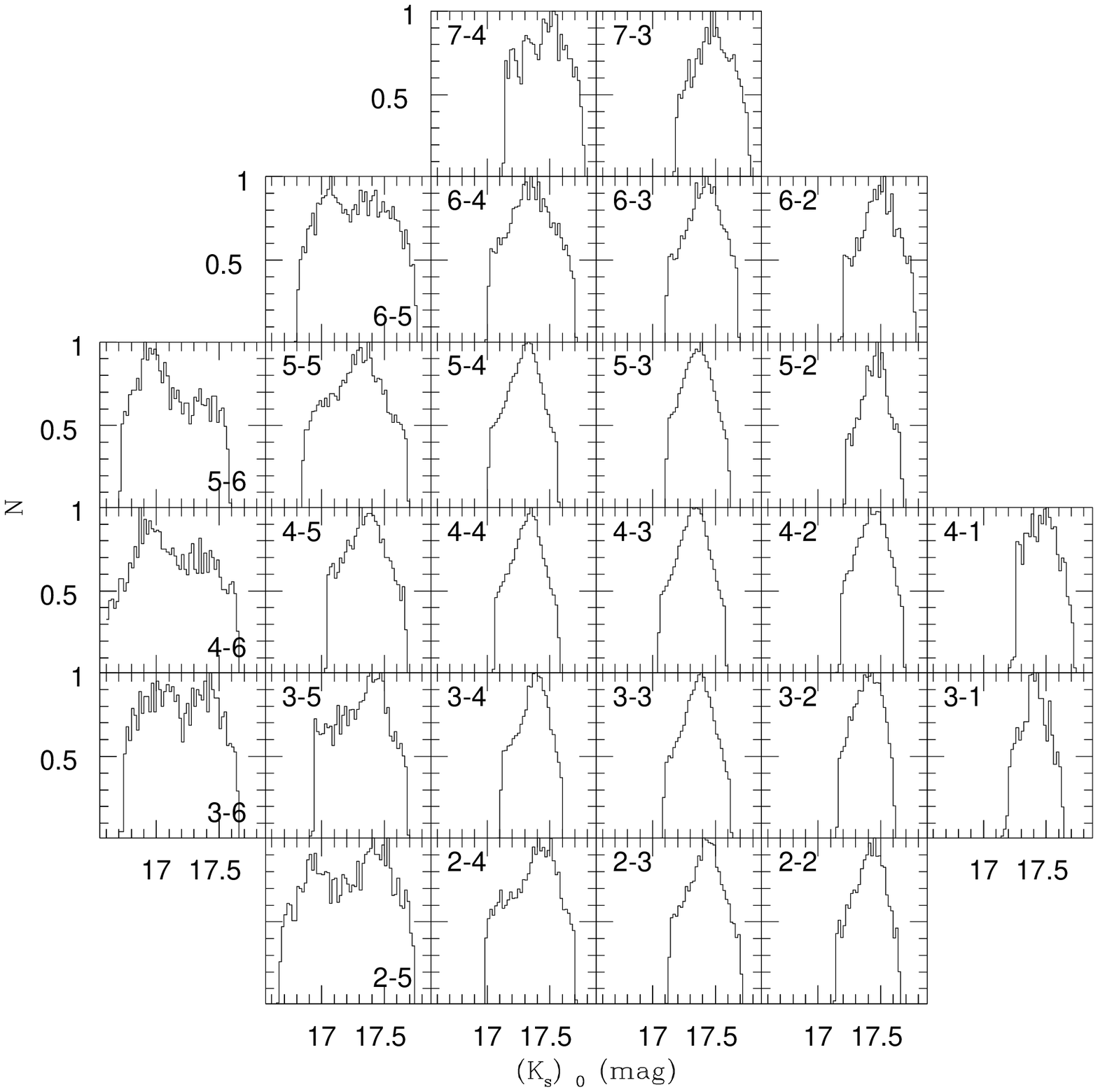}
\caption{Histograms of de-reddened RC star $K_{\rm s}$ magnitudes (bin size of
0.02 mag for $K_{\rm s}=16.5$--18 mag range) for each SMC tile (labelled in the
top--left or bottom--right of each panel). Of particular interest is the
double RC seen in the eastern fields.}
\label{fig:smc:gramd}
\end{figure}

Another consideration is the distribution of distance moduli within the SMC.
This analysis is carried out via histograms of the $K_{\rm s}$ magnitude for
each tile and shown in Figure~\ref{fig:smc:gramd}. It can be seen that for the
eastern periphery (tiles SMC 2\_5, SMC 3\_6, SMC 4\_6, SMC 5\_6 and SMC 6\_5)
the distribution of sources does not only comprise a larger magnitude range
than in the central and western regions but it also is double-peaked. This was
first seen by \citet{subramanian17} in the North--East, but we now see that
its presence is more wide-spread across the East. The fainter peak, at
$K_{\rm s}\sim17.3$--17.4 mag, corresponds to the bulk of the SMC. However, the
brighter peak, at $K_{\rm s}\sim16.9$--17.1 mag, corresponds to the distance to
the LMC; it dominates over the fainter peak in tiles SMC 6\_5, SMC 5\_6 and
SMC 4\_6, which is in the direction of the LMC. Thus it is tempting to link
this component to an effect caused by the LMC.

The north--western-most tiles (SMC 6\_2 and SMC 4\_1) have the dimmest peaks,
around $K_{\rm s}\sim17.5$ mag or $\sim6$ kpc behind the SMC. It is possible
that this is related to the Counter-Bridge \citep[see][]{ripepi17}.

\subsection{Plane fitting}\label{smc:plane}

We fit a plane to the spatial distribution of RC magnitudes, to examine
whether a disc-like component is present in the SMC and to facilitate the
examination of the structure by comparing the RC magnitudes relative to such a
plane of symmetry. Moreover, to find the inclination and position angle (the
orientation) of the SMC a plane must be fit. These values, apart from being of
general interest when comparing with the orientation of the LMC and the path
of motion through the Milky Way halo, may vary with the adopted tracer, thus
revealing possible sub-structure and changes due to the dynamical history of
the system.

In order to fit a plane to the data, we must remove the projection issues
arising from the spherical co-ordinates that our current projection causes.
This is done by transforming the current co-ordinate system of RA, Dec and
measure of distance onto a Cartesian co-ordinate system where the $x$-axis is
anti-parallel to RA, the $y$-axis is parallel to Dec and the $z$-axis points
at the observer. The transformation takes the form of
\citep[e.g.,][]{vandermarelcioni01,subramanian10,subramanian13}:
\begin{equation}
x=-D\times\sin(\,\alpha-\alpha_0\,)\times\cos\delta
\end{equation}
\begin{equation}
y=D\times\sin\delta\times\cos\delta_0
-D\times\sin\delta_0\times\cos(\,\alpha-\alpha_0\,)\times\cos\delta
\end{equation}
\begin{equation}
z=D_0-D\times\sin\delta\times\sin\delta_0
-D\times\cos(\,\alpha-\alpha_0\,)\times\cos\delta,
\end{equation}
where ($\alpha_0$, $\delta_0$), ($\alpha$, $\delta$) and ($D_0$, $D$) are the
RA, Dec and distance of the centre of the system and the sub-region,
respectively.

A least-squares plane fitting procedure finds a solution to the following
equation:
\begin{equation}
z=Ax+By+C.
\end{equation}
The coefficients $A$, $B$ and $C$ are then used to obtain the position angle
(PA) -- of the line of nodes, i.e.\ the intersection of the galaxy plane with
the plane of the sky, where it is measured from North over East
(counter-clockwise) -- and inclination ($i$) as follows:
\begin{equation}
i=\arccos\left(\frac{C}{\sqrt{A^2+B^2+1}}\right)
\end{equation}
\begin{equation}
{\rm PA}=\arctan\left(-\frac{A}{B}\right)+{\rm sign}(B)\times\frac{\pi}{2}
\end{equation}

A single centre was used for the SMC which corresponds to the optical centre:
RA $=13\rlap{.}{\degr}17$, Dec $=-72\rlap{.}{\degr}81$; this is an average of
the values from \citet{devaucouleurs76} and \citet{paturel03}; see also
\citet{degrijs15}. Plane fitting was carried out using all sources, outer
sources (outside the 20\% source density contour level), and the census method
(see Section~\ref{smc:cen}). The `all sources', `outer ' SMC and `census
method' samples contained 561,843 sources, 161,100 sources and 976
sub-regions, respectively. The results are summarised in Table~\ref{tab:smc}.

\begin{table}
\caption{Plane fitting results to RC stars in the SMC.}
\begin{tabular}{ccc}
\hline\hline
\noalign{\smallskip}
 & $i$ ($\degr$)   & PA ($\degr$) \\
\noalign{\smallskip}
\hline
\noalign{\smallskip}
Outer  & $48.5\pm0.1$ & $186.3\pm0.5$ \\
All    & $43.3\pm0.1$ & $179.2\pm0.1$ \\
Census & $35.4\pm1.8$ & $169.8\pm3.1$ \\
\noalign{\smallskip}
\hline
\end{tabular}
\label{tab:smc}
\end{table}

The position angle from the `outer' SMC fit returned a value of
$-173\rlap{.}{\degr}7$ but we use the complementary angle:
$186\rlap{.}{\degr}3$, following the adopted conventions of previous works.
The cause of the rotation for the `outer' fit result being reversed is most
likely due to the centre of the Cartesian system being a void for that fit.

The census method acts as an inner SMC selection due to mainly containing
sub-regions that are within the main body of the SMC. It is expected to
exhibit a shallower inclination because the distribution over distance moduli
is not sharply peaked in the core of the galaxy; also, the most extreme values
are found outside of the main body of the SMC.

That said, the progressively larger inclination as well as rotation of the
position angle, as one moves from the inner part of the SMC towards the
periphery, could be an indication of tidal twisting of the SMC.

\subsection{Deviations from a plane}\label{smc:res}

\begin{figure}
\includegraphics[width=\linewidth,clip=true]{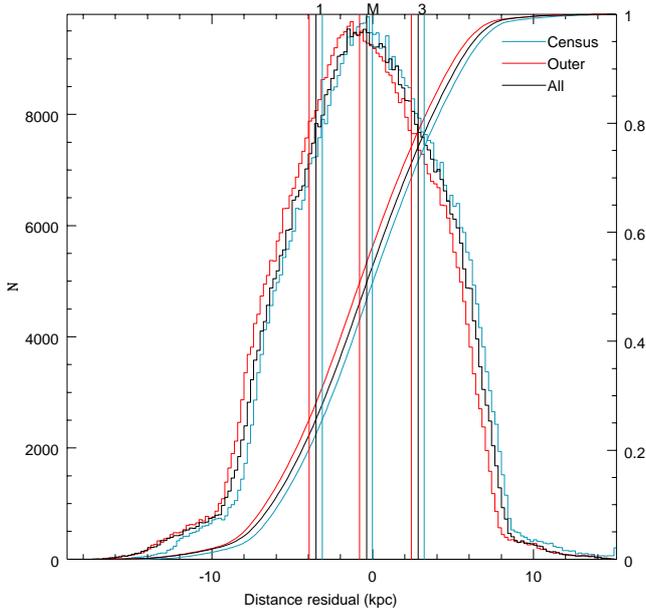}
\caption{Histograms of residuals from the plane fits (bin size of 0.2 kpc) for
`outer' (red), `all' (black) and `census' (turquoise) SMC selections. The
first (leftmost) and fourth (rightmost) quartiles cover a wide range of
distances. Location of the boundaries between the quartiles are indicated with
vertical lines and labelled as `1', `M' and `3', respectively. Cumulative
distributions are overlain (monotonically increasing curves, with values on
the right hand side of the y-axis).}
\label{fig:str:res}
\end{figure}

For each selected RC candidate, the distance can be compared with the distance
calculated at that position on the basis of the fitted plane solution. The
difference between the two values is the `residual'. Figure~\ref{fig:str:res}
shows the histograms of these residuals for the three selections using a bin
size of 0.2 kpc, where negative values lie behind the plane and positive ones
lie in front of the plane.

One can see that the residual range of the two central quartiles is around 6
kpc (between boundaries `1' and `3'), which is a good description of the
line-of-sight depth for the central and western SMC depicted in
Figure~\ref{fig:str:smcdepth} (bottom panel). The central region, in
particular, contains the majority of stars so this is not surprising.
Examining the eastern SMC, line-of-sight depths are double this value. This
was explained by their extended and split RC population as seen in
Figure~\ref{fig:smc:gramd}.

\begin{figure*}
\includegraphics[width=0.49\linewidth,clip=true]{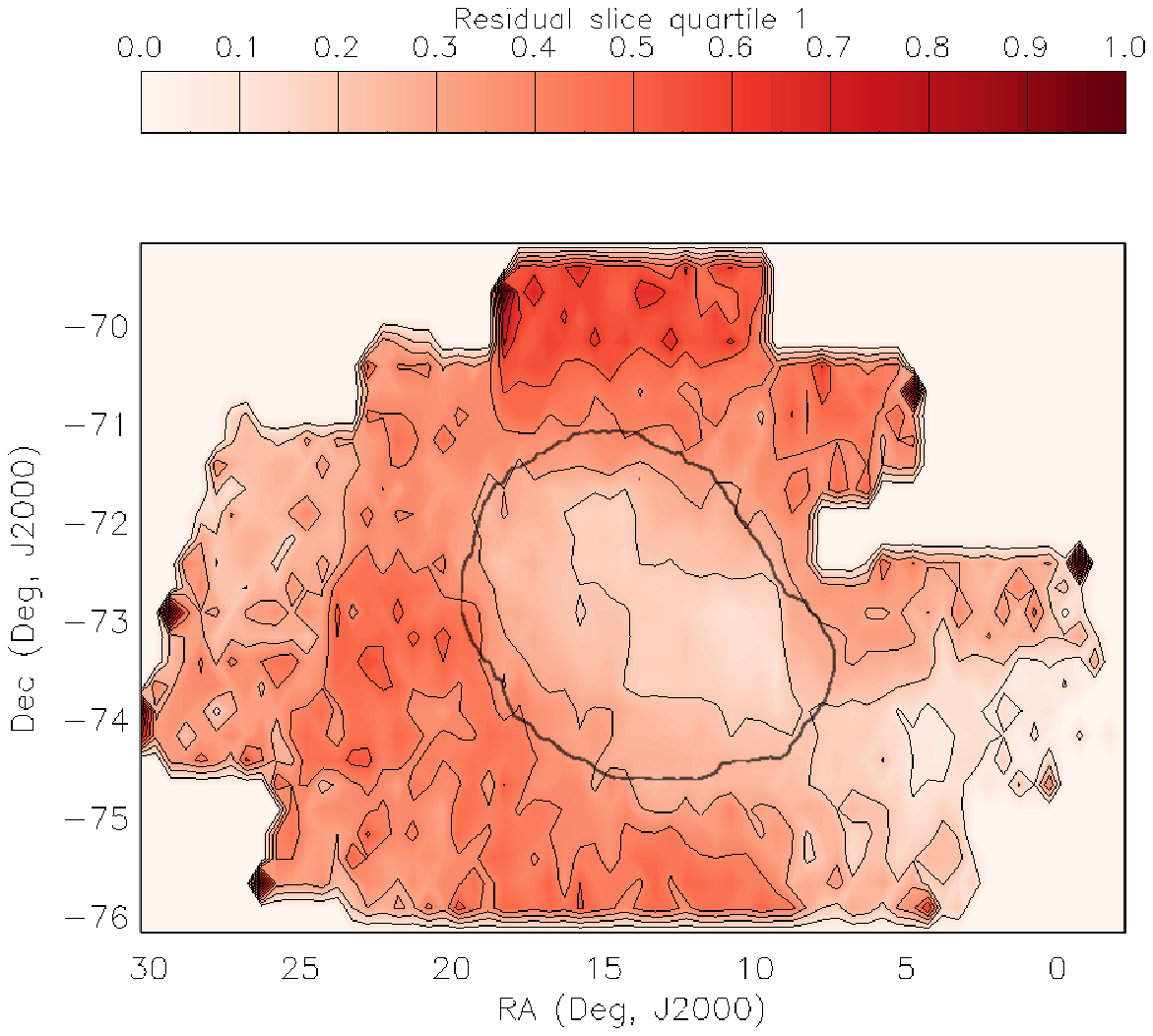}
\includegraphics[width=0.49\linewidth,clip=true]{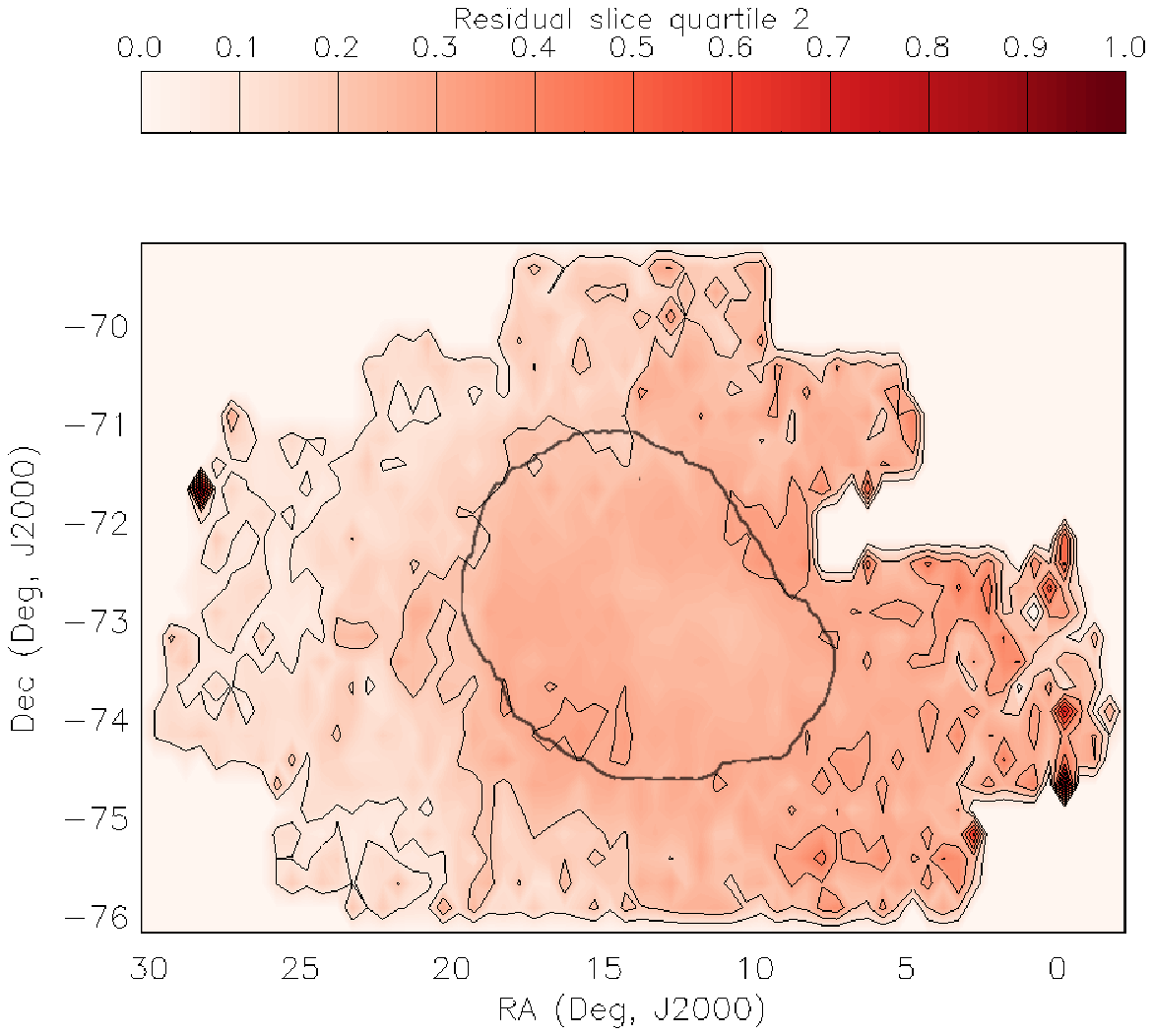}
\includegraphics[width=0.49\linewidth,clip=true]{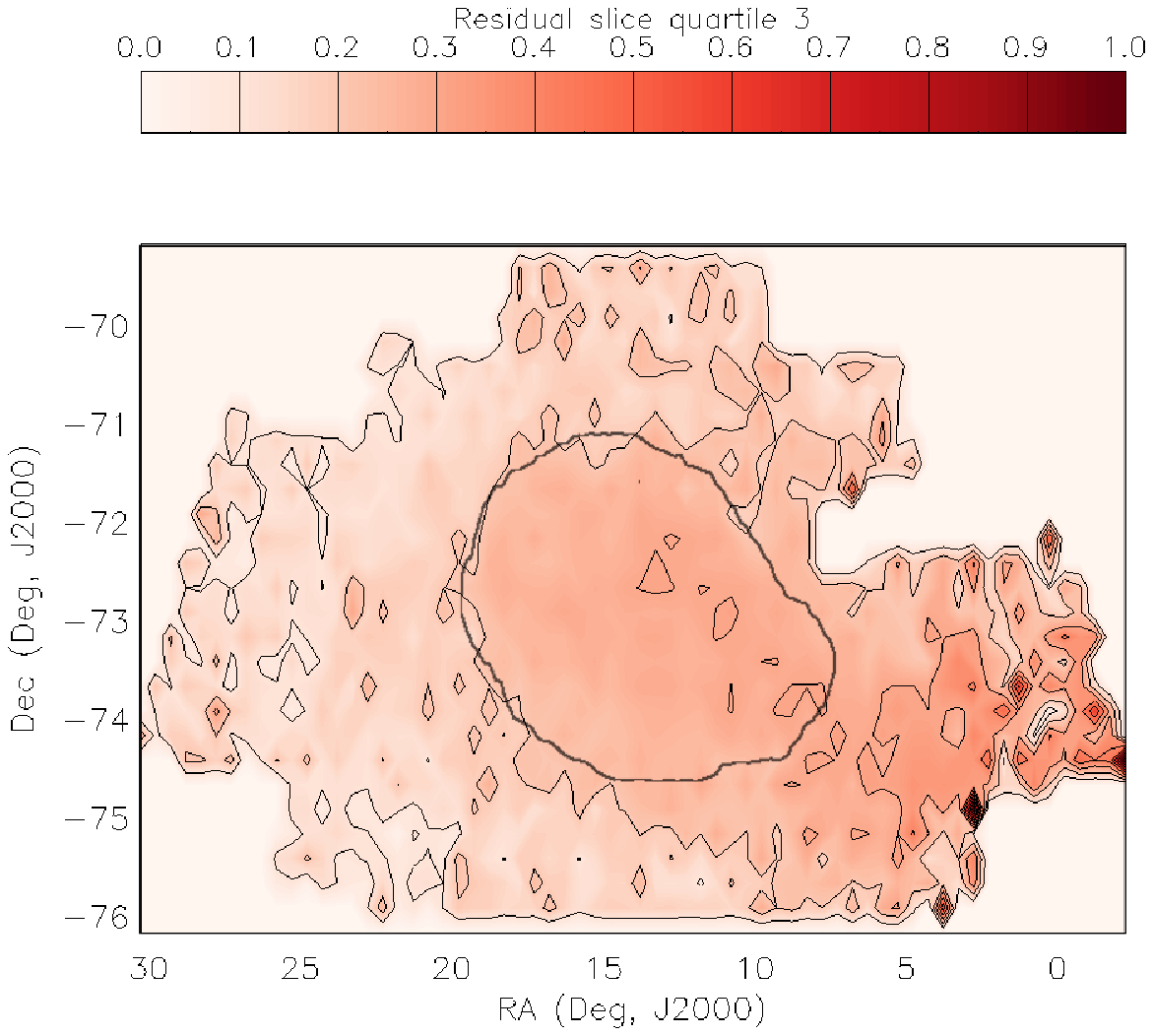}
\includegraphics[width=0.49\linewidth,clip=true]{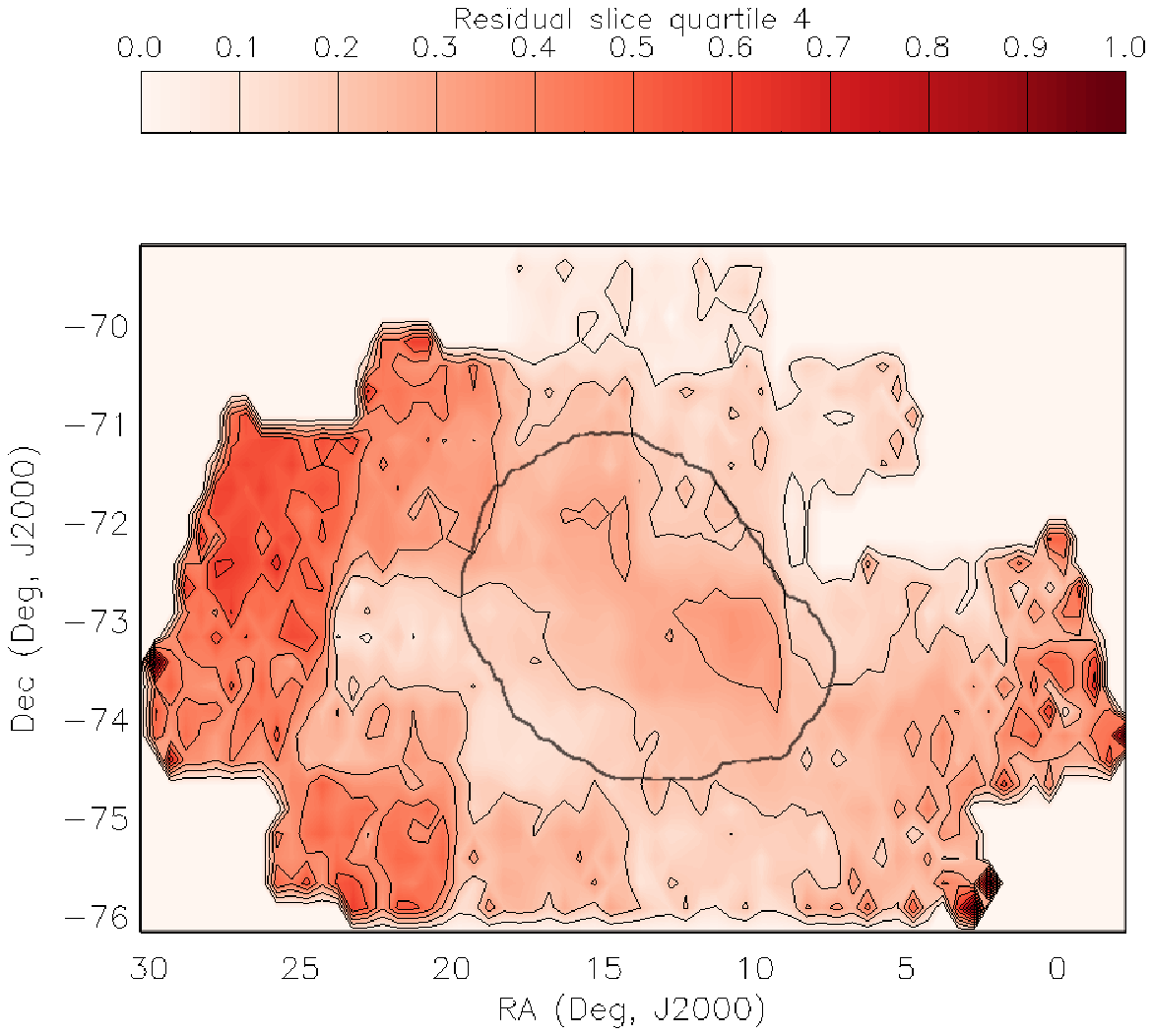}
\caption{Residual slices with respect to a plane fitted to the `all' SMC
selection. Slices are ordered (clockwise) from behind the plane (first
quartile, top--left) to in front of the plane (fourth quartile,
bottom--right). Each slice contains 25\% of the total number of RC stars;
fractions, within bins of size $0\rlap{.}{\degr}5$ in RA and
$0\rlap{.}{\degr}25$ in Dec, can be highly skewed across these distance
slices.}
\label{fig:str:smcsa}
\end{figure*}

We split the residuals into four distance quartiles, from the back to the
front of the plane, shown in Figure~\ref{fig:str:smcsa} for the `all' SMC
plane solution (the other two plane solutions are very similar). The bin size
of these maps is $0\rlap{.}{\degr}5$ in RA and $0\rlap{.}{\degr}25$ in Dec.
Displayed in these contour maps is the fraction of RC stars within a given bin
contained in that quartile. For instance, high fractions in the second and
third quartiles would suggest a compact structure commensurate with the plane
fit, whereas high fractions in the first and fourth quartiles would suggest a
highly extended or even split structure along the line-of-sight.

The main body of the SMC is found mostly in the latter three quartiles whereas
its extent is small in the first (most distant) quartile. This would suggest
that the main body of the SMC is a single component and not too extended in
depth.

The eastern SMC (RA $>25\degr$) is mostly found in the first and last
quartiles though, suggesting some asymmetry and distortion. The eastern tiles
in general also exhibit greater line-of-sight depths (see
Figure~\ref{fig:str:smcdepth}) and bimodal distance distributions (see
Figure~\ref{fig:smc:gramd}), which is reflected here in their presence in the
first and fourth quartiles. Interestingly, in the first quartile there is less
seen in the North--East but more in the South--East and this trend is reversed
in the fourth quartile. With reference to Figure~\ref{fig:smc:gramd}, most
north--eastern tiles (e.g., SMC 5\_6; RA $=25\rlap{.}{\degr}4$, Dec
$=-71\rlap{.}{\degr}6$) have their largest peak of RC stars at relatively
bright magnitudes, i.e.\ closer and thus in front of the plane of symmetry.
The south--eastern tiles (e.g., SMC 3\_5; RA $=21\rlap{.}{\degr}9$, Dec
$=-74\rlap{.}{\degr}0$), on the other hand, contain relatively more fainter RC
stars, which are farther and thus behind the plane of symmetry.

The northern part of the SMC (RA $=10\degr$--$20\degr$, Dec $>-71\degr$) is a
more concentrated example of a single component appearing mainly in the first
(far) quartile but non-existent in the fourth (near) quartile. In contrast the
south--southeastern part (RA $=20\degr$--$26\degr$, Dec $<-75\degr$) is
dominated by the near component.

The westernmost tiles (SMC 3\_1 and SMC 4\_1) are largely absent from the
first quartile. This is confusing considering that the Counter-Bridge
\citep[see][]{ripepi17} is a more distant feature and would be prominent
behind the plane as opposed to in front. It is possible that so close to the
main body of the SMC this distance increase has not yet manifested itself.
Also, foreground contamination is relatively problematic in these sparse SMC
fields.

\section{Discussion}\label{smc:dis}

There are three main topics to discuss based on our new results: the
reddening, the putative disc component and the morphological distortions.\\

\subsection{Reddening (and dust)}\label{smc:dust}

While a comprehensive analysis of the reddening map similar to that in
\citet{tatton13} is beyond te scope of the present study
\citep[cf.][]{bell20}, we will compare the spatial distributions of the
interstellar dust and RC stars.

First, we compare our map with the widely used optical reddening map,
expressed in $E(V-I)$, published by \citet{haschke11}. We limited the
comparison to their map based on RC stars (their Figure 4). Both maps contain
some artificial structure -- small gaps in ours and vertical striping in
theirs. Qualitatively, the agreement is globally satisfactory, with peaks in
similar places along the main body and also around N\,83/84 in the Wing --
though the latter seems more pronounced in their map. Our map covers a larger
area on the sky, however, and clearly displays a smooth, ellipsoidal component
with a reddening gradient away from the main body in all directions. This
component is hard to discern in their map, apart from a global, low level of
$E(V-I)\sim0.04$ mag, corresponding to $A_V\sim0.1$ mag, which seems to fall
off at the extreme south--western and north--eastern edges of their map.
Quantitatively, however, their reddening values correspond to $A_V$ values
that are about three times lower than ours. A similar discrepancy had been
noted for the 30\,Doradus field in the LMC \citep{tatton13}, which was
attributed to a combination of reddening bias and contamination from non-RC
stars resulting from the more simplistic RC selection criteria adopted by
\citet{haschke11}. \citet{gorski20} improved upon the work by Haschke et al.;
their mean and maximum reddening values correspond to $A_V\sim0.25$ and
$\sim0.70$, respectively, which match ours within a factor two.

Second, we compare our map with other ISM tracers. In the H$\alpha$ map of
\citet{kennicutt95} the western side of the main body has similar emission
levels to the south--eastern side. In the H\,{\sc i} map of
\citet{stanimirovic99} and the mid- and far-IR maps of \citet{bolatto07} and
\citet{leroy07}, on the other hand, the western side of the main body has
weaker emission than the eastern. The differences between the H$\alpha$ study
and the other studies suggest there are changes in the ionisation state -- and
possibly the dust-to-gas ratio -- between the western and eastern parts of the
SMC main body. Dust emission is dependent on dust temperature and therefore
the interstellar radiation field, whereas extinction is not. Thus, extinction
is a more direct probe of dust.

In our extinction map weaker extinction is seen in the north--western side of
the main body than in the South--East (see also Figure~\ref{fig:smc:cen},
top--right panel), corroborating what is seen in the H\,{\sc i} and
IR-emission maps and the reddening inferred using background galaxies
\citep{bell20}. The strongest extinction features seen are best traced by the
8-$\mu$m data of \citet{bolatto07}, suggesting that near-IR reddening may
trace predominantly small grains (which are easily heated and thus shine
brightly at relatively short IR wavelengths). The 160-$\mu$m data of
\citet{gordon09} cover the Wing and main body and, unlike the previously
mentioned maps, exhibit strands of emission connecting the main body with the
Wing. In contrast, our map shows a more even distribution of moderate
extinction between the two and strands are not apparent. This discrepancy can
be explained by recalling the drawback of using extinction, which is that it
is only measured in front of stars, whereas the dust emission is seen wherever
it is located. Perhaps these strands of emission are predominantly located
behind the stars resulting in less extinction measured than dust emission
observed. Alternatively, if the strands are very thin, then few sightlines
would cross them.

\begin{figure}
\includegraphics[angle=0,clip=true, width=\linewidth]{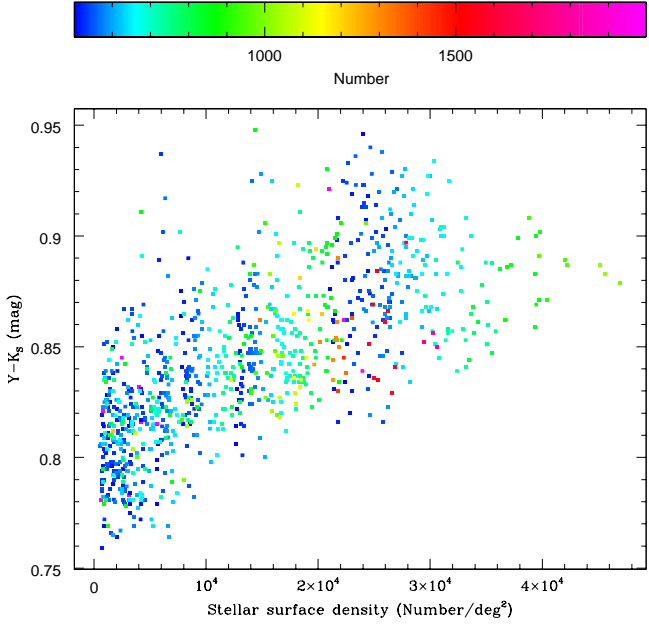}
\caption{$Y-K_{\rm s}$ colour vs.\ regional census sub-region density of RC
stars, showing a positive correlation (albeit with considerable scatter). The
colour scale quantifies the number of sources per sub-region.}
\label{fig:smc:cenden}
\end{figure}

Another interpretation of our data comes from looking at the $Y-K_{\rm s}$
colours for the regional census sub-regions (Figure~\ref{fig:smc:cenden}). A
trend can be seen of average extinction increasing with respect to source
density (i.e.\ the inner SMC has higher extinction than the outer SMC). This
means that, to some extent, stars and dust are well mixed and follow the same
potential. The scatter around this correlation is probably intrinsic, and the
fact that the densest regions are not located where the colours reach the
reddest values probably traces the recent star formation occurring in stripped
gas devoid of an old stellar population. A particular high-extinction feature
can be seen in the census map (Figure~\ref{fig:smc:cen}, top--right panel)
around RA $=18\rlap{.}{\degr}5$, Dec $=-73\rlap{.}{\degr}3$. This is
associated with the molecular cloud complex N\,83/84.

\begin{figure}
\includegraphics[angle=0,clip=true, width=\linewidth]{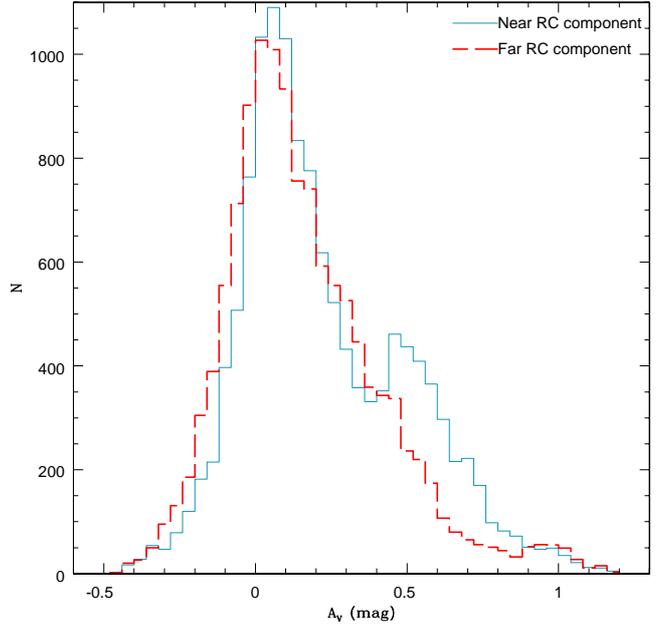}
\caption{Histogram (bin size of 0.04 mag) comparing the extinction
distribution of RC stars in the near-Wing component (turquoise, solid) to that
in the far-Wing component (red, dashed).}
\label{fig:smc:wingcomp}
\end{figure}

Finally, we examine the extinction distribution of five Wing tiles (SMC 2\_5,
SMC 3\_6, SMC 4\_6, SMC 5\_6 and SMC 6\_5) by splitting their populations into
a near component ($K_{\rm s}<17.2$ mag, 12,855 RC stars) and a far component
($K_{\rm s}>17.2$ mag, 12,339 RC stars). The resulting histogram is shown in
Figure~\ref{fig:smc:wingcomp}. We might have expected more distant stars to
experience higher extinction than stars in front, yet this is not seen. The
overall distributions are in fact very similar, and no systematic shift is
discernible. A feature that stands out more clearly is an excess of nearer
stars apparently experiencing $A_V\sim0.5$--0.8 mag. This can be traced back
to the RGB bump, in Fig.~\ref{fig:selection} at $K_{\rm s}\approx 17.0$ mag and
$(Y-K_{\rm s})\approx 0.95$ mag, and is seen in all tiles (see
Section~\ref{smc:pop} for a discussion of population effects). The far
component is fainter than the RGB bump, so is not affected by it, but the RGB
itself still leaves an imprint. As Fig.~\ref{fig:selection} shows, this part
of the RGB is less populated than the RGB bump, closer to the RC but bent more
and thus the effect is both weaker and more wide-spread over $A_V$ -- and
therefore only appears as an inconspicuous shoulder to the extinction
distribution around $A_V\sim0.4$ mag. The reddening map is not dramatically
affected by this inherent feature of the methodology, but it is important to
be aware of its existence when using the map in all its fine detail.

\subsection{Disc component}

The inclination and position angle values found in the literature (see
Table~\ref{tab:smcpai}) greatly vary depending on the adopted tracer and
method. This makes a direct comparison tricky, but potentially insightful.
Younger Cepheid populations tend to have very steep inclinations -- even
though they do not strictly speaking adhere to a flat disc geometry
\citep[][their Fig.\ 17]{ripepi17}. Older populations, on the other hand, tend
to have very shallow, almost non-existent inclination -- the oldest
populations are more likely to be more spheroidal, as in the Milky Way halo,
rather than being represented by a disc-like structure. One would therefore
expect intermediate-aged populations traced, for instance, by RC stars to be
somewhere between those. This is exactly what we found.

\citet{subramanian12} also used RC stars but found a very small inclination,
$i=0\rlap{.}{\degr}5$. Their methodology however was to use axes ratios and to
limit the sample to a small depth range, which may have inadvertently
introduced a bias. The position angles we found are nearest to those
determined by \citet{subramanian15}, who used Cepheids, and \citet{rubele15},
who fitted entire CMDs and found a mean $i=39\degr$.

In conclusion, while the SMC does not exhibit an obvious disc component there
is a clear trend for the system to become more inclined with respect to the
plane of the sky the younger the (stellar) tracer is. Since the RC stars
already show this progression this means that the dynamical events that led to
this time-dependent morphology must have occurred on Gyr timescales. Indeed,
while most of the stellar mass in the SMC was generated $>4$ Gyr ago
\citep{sabbi09,rezaeikh14,rubele18} there is evidence in the SFH of the SMC of
a strong interaction with the LMC to have occurred $\sim2$ Gyr ago
\citep{harris04,noel09,rezaeikh14,rubele15}. A more recent interaction
$\sim0.2$ Gyr ago coincident with the closest approach to the Milky Way, also
seen in the SFH \citep{harris04,noel09,sabbi09,rubele15}, may have caused
further distortion of the gaseous disc, from which the younger population
traced by the Cepheids formed.

\subsection{Distortions and other structures}
\label{distortions}

The outer regions of the SMC have been transformed by interactions leading to
stripping and spatially stretched populations. \citet{dias16} described the
SMC as being composed of three bodies: the eastern Wing, the central main
body, and the western halo. We find the intermediate-age main body traced by
RC stars to be structurally simple and symmetrical, consistent with a
spheroidal component. The eastern Wing, on the other hand, is split into two
distinct populations (both traced by intermediate-age RC stars), of which the
brighter one clearly lies in front of the main body (Fig.~\ref{fig:str:smcsa},
bottom--right). We have been able to comprehensively trace these two
components across the VMC footprint, allowing us to relate them to what other
observational and theoretical studies have found. We find a fourth structure,
in the North (Fig.~\ref{fig:str:smcsa} top--left); this has previously been
described as an extension to the Wing \citep[e.g.,][]{gardiner91}, but
\citet{sun18} showed this area to contain a shell of young stars which is
clearly separated from the general Wing area. The intermediate-age population
of RC stars in the North need not be related to these young stars, though. We
compare our results with previous analyses of Cepheids, RR\,Lyr{\ae} and RC
stars, in light of these structures.

\citet{ripepi17} showed that both the young and old Cepheids (old Cepheids are
$\sim300$ Myr old, i.e.\ still much younger than RC stars) populate off-centre
structures in the direction of the LMC. \citet{nidever13a} found that the
eastern side of the SMC has a bimodal distribution with a near component at 55
kpc and a far component at 67 kpc. In the same eastern regions we see a large
line-of-sight depth but the sources are located mainly in front of the plane
(i.e.\ a near component). However, these features may not be representative of
the full content of the entire Wing. Young structures such as NGC\,602 (RA
$=22\degr$, Dec $=-73\rlap{.}{\degr}6$) and N\,83/84 (RA
$=18\rlap{.}{\degr}5$, Dec $=-73\rlap{.}{\degr}2$) trace sub-structure not
seen in the intermediate-age populations. \citet{hammer15} showed that past
interactions between the LMC and SMC may have resulted in multiple tidal and
ram-pressure stripped structures; while tidal effects are largely
non-discriminatory in age, ram pressure can spatially separate gas in which
new stars may form, from pre-existing (i.e.\ older) stellar populations.

\citet{subramanian12} selected optical photometry of RC stars and RR\,Lyr{\ae}
(which are older than RC stars by a factor of a few, {\it viz.}\ $>10$ Gyr)
and found typical line-of-sight depths of $4.6\pm1.0$ kpc which increased to
6--8 kpc in the North--East. \citet{muraveva18} used VMC photometry of
RR\,Lyr{\ae} and found an ellipsoidal distribution. Line-of-sight depths were
in the range of 1.5--10 kpc, with an average of $4.4\pm1.0$ kpc; these are
around a factor of two smaller than the values derived here for RC stars (for
the central quartiles we see a range of 0.5--20 kpc with an average of 7.8
kpc). We confirm the larger line-of-sight depth in the North--East; an
increased depth in the South--West is less obvious and easier to see in the
full line-of-sight depth map (Fig.~\ref{fig:str:smcdepth}).

Using a limited VMC data set, \citet{subramanian17} found a double RC feature
manifesting itself as a distance bimodality in the eastern SMC with a
`foreground' population $\sim11$ kpc nearer to us than the main SMC body. We
confirmed this bimodality (see Figure~\ref{fig:smc:gramd}) and showed it is
also seen within adjacent tiles not studied by \citet{subramanian17}: tiles
SMC 2\_5 (RA $=22\rlap{.}{\degr}5$, Dec $=-75\rlap{.}{\degr}1$), SMC 4\_6 and
SMC 3\_6 (RA $=27\degr$, Dec $=-72\degr$ to $-74\degr$). The extent of the RC
bimodality has since been traced further by El Youssoufi et al.\ (submitted).
In addition, the northern tile SMC 7\_4 (RA $=15\rlap{.}{\degr}8$, Dec
$=-69\rlap{.}{\degr}9$) also exhibits some bimodality but spanning a less
extended magnitude range. The closer of the two components is found around
$K_{\rm s}\simeq17.0$ mag making it much nearer to the LMC ($K_{\rm s}\simeq16.9$
mag) than to the SMC ($K_{\rm s}\simeq17.4$ mag).

The RR\,Lyr{\ae} do not show a distance bimodality \citep{muraveva18} -- while
the eastern part of the SMC seems tilted towards us (their figure 15), there
is a distinct lack of RR\,Lyr{\ae} $\sim0.5$ mag brighter than those in the
main body of the SMC (see their figure 8). This might suggest that the nearer
RC component resulted from ram-pressure stripping of gas a few Gyr ago which
led to subsequent formation of stars including what are now RC stars and
Cepheids but without the presence of older stars such as RR\,Lyr{\ae}. A tidal
origin would have affected stars of all ages similarly. This could mean that
the Wing is, in origin, an intermediate-age structure that has persisted for a
few Gyr; more recent tidal interactions could have shaped it further still.

An origin for the Bridge and the nearer RC component, and possibly the Wing,
much earlier than a few hundred Myr is also consistent with the lower
metallicity in the gas from which younger stars have formed in western parts
of the Bridge \citep{rolleston99,lehner08,gordon09}. If these structures had
formed from the main body of the SMC only a few hundred Myr ago then their
metallicity ought to be indistinguishable from that of the ISM and young stars
within the main SMC body. However, the metallicity in the Wing is much more
similar to that within the main SMC body \citep{lee05}. This may be explained
if, since the initial dislodging from the main body a few Gyr ago, star
formation and chemical evolution within the eastern elongated structures
continued on a par with that in the main body of the SMC in the densest parts
closest to the SMC (Wing), whilst it proceeded much more slowly in the more
tenuous parts further away (Bridge). Direct determination of the metallicity
of the nearer RC stars could establish the connection between this
intermediate-age component and the younger stars in the Wing, or the older
stars in the Bridge.

While recent hydrodynamical simulations by \citet{Wang19} show the separation
of gas and stars due to the differing effects from tidal disruption and ram
pressure interaction, they fail to reproduce the split RC and in particular
the location on the sky of the far component. It is possible that this is a
function of the exact (pre-)collision characteristics $>200$ Myr ago.

It is not just the East that is expected to display a distorted morphology. A
Counter-Bridge originating in the West is predicted by the models of
\citet{diaz12} as a remnant of the last close LMC--SMC encounter. However, due
to the distances involved (70--80 kpc) its existence has so far only been
suggested by \citet{ripepi17} and \citet{niederhofer18}. We may be seeing a
somewhat more distant RC clump in the far West (Fig.~\ref{fig:smc:gramd}, tile
SMC 4\_1), though we caution that the source density is low. However, the
Counter-Bridge is predicted to wrap around the SMC, from the South--West to
the North--East. The dominance of `distant' RC stars in the North
(Fig.~\ref{fig:str:smcsa}, top--left) could plausibly be associated with that
part of the Counter-Bridge. Our finding is corroborated by El Youssoufi et
al.\ (submitted). In fact, the foreground component appears to cross the main
body of the SMC from North--East to South--West (Fig.~\ref{fig:str:smcsa},
bottom--right); while this is dependent on the location of the `plane' that
was fit, the absence of a background component at the location of the main
body of the SMC seems to confirm this interpretation. This suggests that both
the Wing/Bridge and Counter-Bridge may have been stripped from the SMC first,
trailing the SMC in its wake, before having been stretched by tidal effects.

\section{Conclusions}\label{smc:conc}

This work used near-IR survey data of the SMC from the recently completed VMC
survey, and reliable, homogenised PSF photometry covering the entire galaxy,
to map its structure as traced by intermediate-age RC stars in unprecedented
detail. The near-IR has the advantage that the effects from interstellar
reddening and population age and metallicity are much reduced compared with
the optical. That said, we did derive also a reddening map, which showed
general agreement with dust emission and which was used to correct the
$K_{\rm s}$-band magnitudes of the RC stars.

The large line-of-sight depths seen in the East led us to re-examine the
magnitude distributions of the RC stars across the SMC. From this, a double RC
component was seen whose nearer component is around the LMC distance. While
this feature had been reported in \citet{subramanian17}, here we expanded upon
that study by showing how this feature varies across a larger area of the sky.

The residuals from fitting a plane once again emphasised that these eastern
regions reveal structure that is not described by a plane in the periphery of
the SMC. The inclination derived from plane fitting solutions
($i=35\rlap{.}^\circ4\pm1\rlap{.}^\circ8$ for the inner-dominated census
selection) falls between the very shallow inclinations reported for older
RR\,Lyr{\ae} populations ($i\simeq2\degr$) and younger Cepheid populations
($i\simeq60\degr$). A distant population of RC stars seen in the North may be
associated with the Counter-Bridge, as predicted by tidal interaction models.

The structures within the SMC mainly arise from its past close interactions
with the LMC, which is evident in the sense that the East -- which is nearest
to the LMC -- has the greatest distortions. The fact that some of these
structures are traced by RC stars but not RR\,Lyr{\ae} stars means that tidal
effects cannot fully account for them and star formation a few Gyr ago in
ram-pressure stripped gas must be invoked.

\section*{Acknowledgements}

We thank the referee for their constructive report which helped improve this
paper. We thank Jim Emerson and L\'eo Girardi for comments on an earlier
version of the manuscript. BLT acknowledges support from an STFC studentship
at Keele University during which most of this work was accomplished, and from
a Keele University visitorship to finalise its publication. University of
Hertfordshire's computing resources were used to produce the PSF photometry.
MRC and CPMB acknowledge support from the European Research Council (ERC)
under the European Union's Horizon 2020 research and innovation program (grant
agreement No.\ 682115). SS acknowledges support from the Science and
Engineering Research Board, India through a Ramanujan Fellowship. We thank the
Cambridge Astronomy Survey Unit (CASU) and the Wide Field Astronomy Unit
(WFAU) in Edinburgh for providing calibrated data products under the support
of the Science and Technology Facility Council (STFC) in the UK. The data used
for this study are based on observations collected at the European
Organisation for Astronomical Research in the Southern Hemisphere under ESO
programme 179.B-2003.

\section*{Data Availability}
The photometry data used in this paper are available in the VISTA Science Archive (VSA), at \url{http://horus.roe.ac.uk/vsa} \\
The reddening map catalogues derived from this data (presented in Figure 4) will be available in Centre de Donnees astronomiques de Strasbourg (CDS), at \url{https://cds.u-strasbg.fr}

\bibliographystyle{mnras}
\bibliography{all}
\label{lastpage}
\end{document}